\documentclass[twocolumn]{29onr_art}
\usepackage{natbib}
\usepackage{ulem}
\usepackage{graphicx}
\usepackage{fixltx2e}
\usepackage{amsmath,float}
\usepackage{graphicx}
\usepackage[english]{babel}
\usepackage{epsfig}
\usepackage{color}
\usepackage{epstopdf}
\usepackage{dblfloatfix}
\usepackage[colorlinks=true,
            urlcolor=blue,
            linkcolor=docgreen,
            citecolor=docgreen,
            bookmarks=true,
            pdftitle={The Turbulent Boundary Layer on a Surface-Piercing Vertical Wall },
            pdfauthor={Nathan Washuta},
            pdfsubject={Paper submitted for the 30th Naval Hydrodynamics Symposium},
            pdfkeywords={Boundary layer, free surface},
	    bookmarksopen=false,
            pdfpagemode=None]{hyperref}

\definecolor{docgreen}{rgb}{0,.5,0}

\pdfoutput = 1

\begin{document}

\title{The Turbulent Boundary Layer on a Horizontally Moving, Partially Submerged, Surface-Piercing Vertical Wall}

\author{
Nathan Washuta, Naeem Masnadi, and James H. Duncan}

\affiliation{\small University of Maryland, College Park, Maryland, U. S. A.}

\maketitle

\begin{abstract}

The complex interactions between turbulence and the free surface, including air entrainment processes, in boundary layer shear flows created by vertical surface-piercing plates are considered. A laboratory-scale device was built that utilizes a surface-piercing stainless steel belt that travels in a loop around two vertical rollers, with one length of the belt between the rollers acting as a horizontally-moving flat wall. The belt is operated both as a suddenly-started plate to reproduce boundary layer flow or at steady state in the presence of a stationary flat plate positioned parallel to the belt to create a Couette flow with a free surface. Surface profiles are measured with a cinematic laser-induced fluorescence system in both experiments and air entrainment events and bubble motions are observed with stereo underwater white-light movies in the suddenly started belt experiment. It is found that the RMS surface height fluctuations, $\eta$, peak near the boundaries of the flows and increase approximately linearly with belt speed. In the Couette flow experiments, a dominant peak in the spectrum of $\partial\eta/\partial t$ is found at the dimensionless frequency $fH/U\approx 0.7$. In the suddenly started belt experiment, surface fluctuations appear to be strongly influenced by sub-surface turbulence within the boundary layer, while ripples propagate freely when farther from the belt.  Additionally, some mechanisms for air entrainment are observed and comparisons are made to predictions of incipient entrainment conditions in \cite{broc:2001}.

\end{abstract}

\section{1. Introduction}

\begin{figure}
\begin{center}
\includegraphics[trim=0 0.1in 0 0.1in, clip=true,scale=0.13]{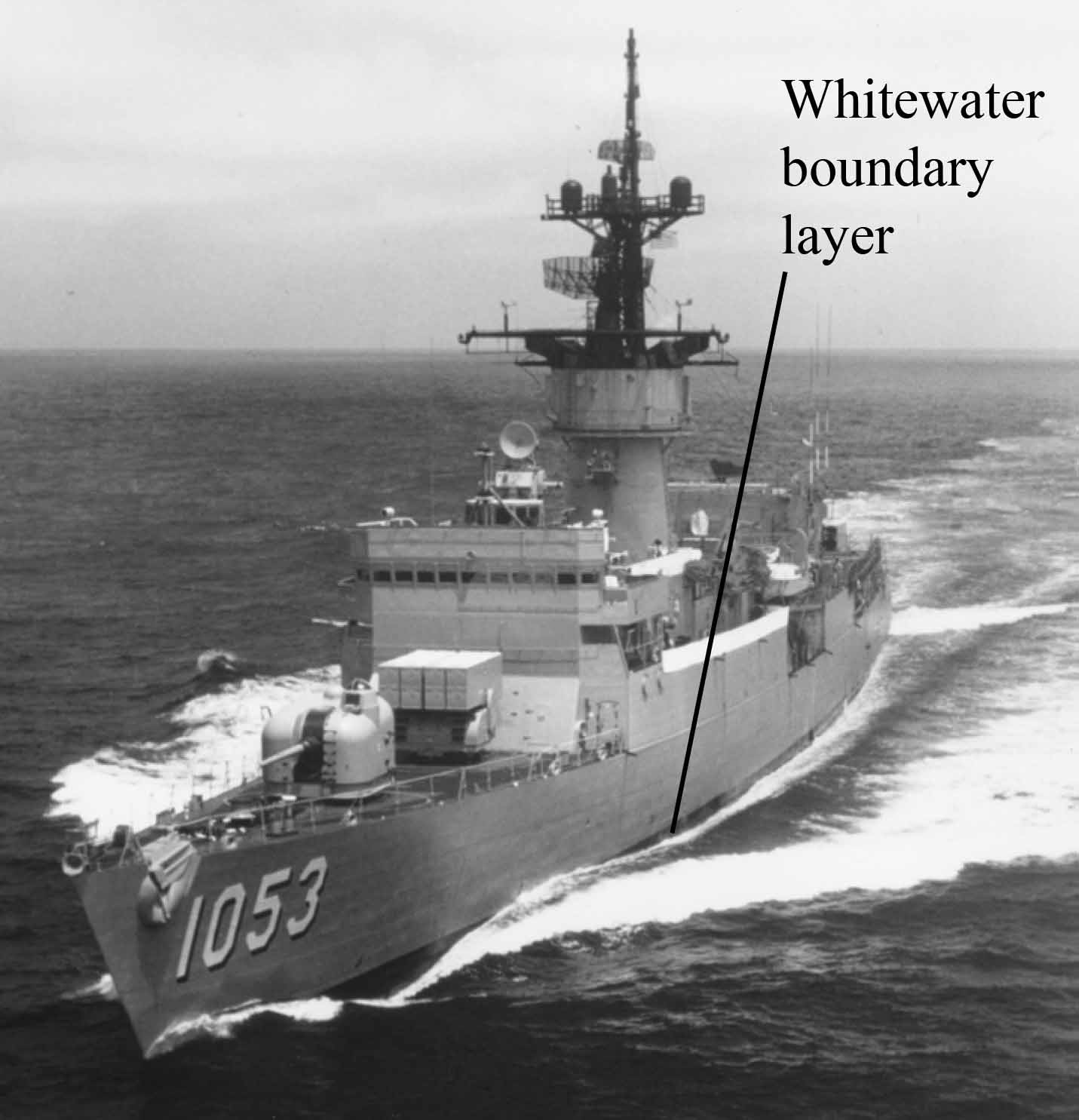}
\end{center}
\vspace*{-0.17in} \caption{Photograph of naval combatant ship showing zone of white water next to the hull.} \label{fig:ship}
\end{figure}

Turbulent boundary layers near the free surface along ship hulls and surface-piercing flat plates have been explored by a number of authors, see for example \cite{Longo1998}, \cite{Sreedhar1998}, \cite{Stern1989} and \cite{Stern1993}.  However, even though it has long been observed that there is a layer of white water next to the hulls of naval combatant ships moving at high speed, see for example the photograph in figure~\ref{fig:ship}, the entrainment of air at the free surface in ship boundary layers has received relatively little attention. It is not known whether this white water is the result of active spray generation and air entrainment due to turbulence in the boundary layer along the ship hull or the result of spray and air bubbles that are generated upstream in the breaking bow wave and then swept downstream with the flow. In the free surface boundary layer, the air entrainment process is controlled by the ratios of the turbulent kinetic energy to the gravitational potential energy and the turbulent kinetic energy to the surface tension energy. The ratio of the turbulent kinetic energy to the gravitational potential energy is given by the square of the turbulent Froude number ($Fr^2 = q^2 / (g L)$) and the ratio of turbulent kinetic energy to surface tension energy is given by the Weber number ($We = \rho q^2 L/ \sigma$), where $g$ is the acceleration of gravity, $\rho$ is the density of water, $\sigma$  is the surface tension of water, $q$ is the characteristic magnitude of the turbulent velocity fluctuations and $L$ is the length scale of this turbulence. 

\begin{figure}
\begin{center}
\includegraphics[width=3in]{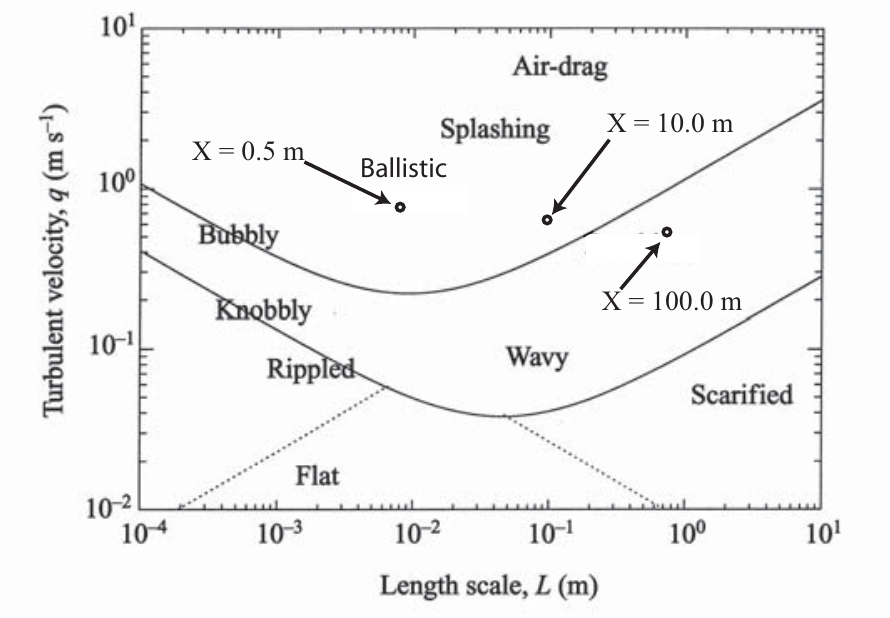}
\end{center}
\vspace*{-0.2in} \caption{Regions of various types of surface motions for free surface turbulence with velocity fluctuation magnitude $q$ (vertical axis) and length scale $L$ (horizontal axis), from \cite{broc:2001}. Air entrainment and spray production occur in the upper region, above the uppermost curved line. The three data points are values obtained for the turbulent boundary layer on a flat plate with $q$ taken as the rms of the spanwise (which is vertical for the boundary layer along a ship hull) velocity fluctuation ($w'$) and L taken as the boundary layer thickness ($\delta$).} \label{fig:brocchiniperegrine}
\end{figure}

Several authors have applied theory and numerical methods to explore the interaction of turbulence and a free surface, see for example \cite{Shen2001}, \cite{Guo2009}, \cite{kim:2013} and \cite{broc:2001}.  
\cite{broc:2001} have used scaling arguments to predict the critical Froude and Weber numbers above which air entrainment and spray generation will occur due to strong free-surface turbulence. Figure~\ref{fig:brocchiniperegrine}, which is from their paper, shows the boundaries of various types of surface undulations on a plot of $q$ versus $L$. The upper region of the plot is the region of air entrainment and droplet generation. We have used classical boundary layer correlations to make estimates of $q$ (taken as the root-mean-square vertical component of the turbulent velocity fluctuations) and $L$ (taken as the boundary layer thickness) at three streamwise positions in a ship boundary layer and plotted these points on the $q$-$L$ map in figure~\ref{fig:brocchiniperegrine}. As can be seen from the figure, the points are clearly in the air entrainment region of the plot, especially the points near the bow. Thus, air entrainment due to strong turbulent fluid motions in the hull boundary layer at the free surface is a likely cause of the layer of white water. 

The difficulty with laboratory experiments on bubble entrainment and spray stems from the fact that the experiments are performed in the same gravitational field as found in ship flows and that the only practical liquid available is water, as is also found in the ocean. Thus, with $g$, $\rho$ , and $\sigma$ the same in the field and in the laboratory, one must attempt to achieve full-scale flow speeds in order to obtain Froude and Reynolds similarity with field conditions. Also, even if full scale-values of $q$ and $L$ were obtained by towing a surface piercing flat plat with the length of the ship at high speed in a ship model basin, the free surface flow would include a bow wave which would obfuscate the source of the bubbles and spray. Another problem is that in order to obtain realistic entrainment/spray conditions and bubble/droplet size distributions, these experiments should be performed in salt water which is not typically used in ship model basins.

In view of the above difficulties in simulating air entrainment due to the turbulent boundary layer, we have built a novel device  that produces an approximation of a full scale ship boundary layer in the laboratory. This device, called the Ship Boundary Layer Simulator (SBLS) generates  a temporally evolving boundary layer on a surface piercing flat surface. The surface consists  of a stainless steel belt loop that is 1.0~m wide and about 15 m long. The belt is mounted on two vertically oriented drums  as shown in figure~\ref{fig:TankSchem}. The drums are driven by hydraulic motors  and the entire device is placed in a large open-surface water tank as shown in the figure. Before each experimental run, the belt and the water in the tank are stationary. The water level is set just below the top edge of the belt and the flow outside the belt loop on one of the long lengths between the rollers is studied. The belt speed can be set to values between 2.5 and 15~m/s. Uncertainty in measured belt speed varies depending on the desired velocity, but typically remains below 5 percent for the conditions presented below.

Two experiments are discussed in this paper.  In the first experiment, a Couette flow with a free surface is created by placing a fixed long surface-piercing flat plate parallel to and about 4~cm away from the belt test length.  In this case, measurements of the free surface height fluctuations with the belt running in a steady state conditions are performed.  In the second experiment, the belt is accelerated from rest with a 3$g$ acceleration until it reaches a pre-defined speed which is held steady for a short time. The flow on the surface of the belt in this case is a simulation of the flow seen by a stationary observer in the ocean as a ship, that makes no waves, passes by at constant speed. The distance $x$ along the ship hull corresponding to any time $t$ after the belt begins to move is essentially the distance traveled by the belt, $x = Ut$, where $U$ is the ship/belt speed. 

The remainder of this paper is divided into three sections.  The experimental setup is described in the following section 2.  The results and discussion are given in Section 3 and the conclusions of this study are presented in Section 4.

\section{2. Experimental Details}

\begin{figure*}[!htb]
\begin{center}
\includegraphics[trim=0 2.5in 0 0.00in,clip=true,scale=0.5]{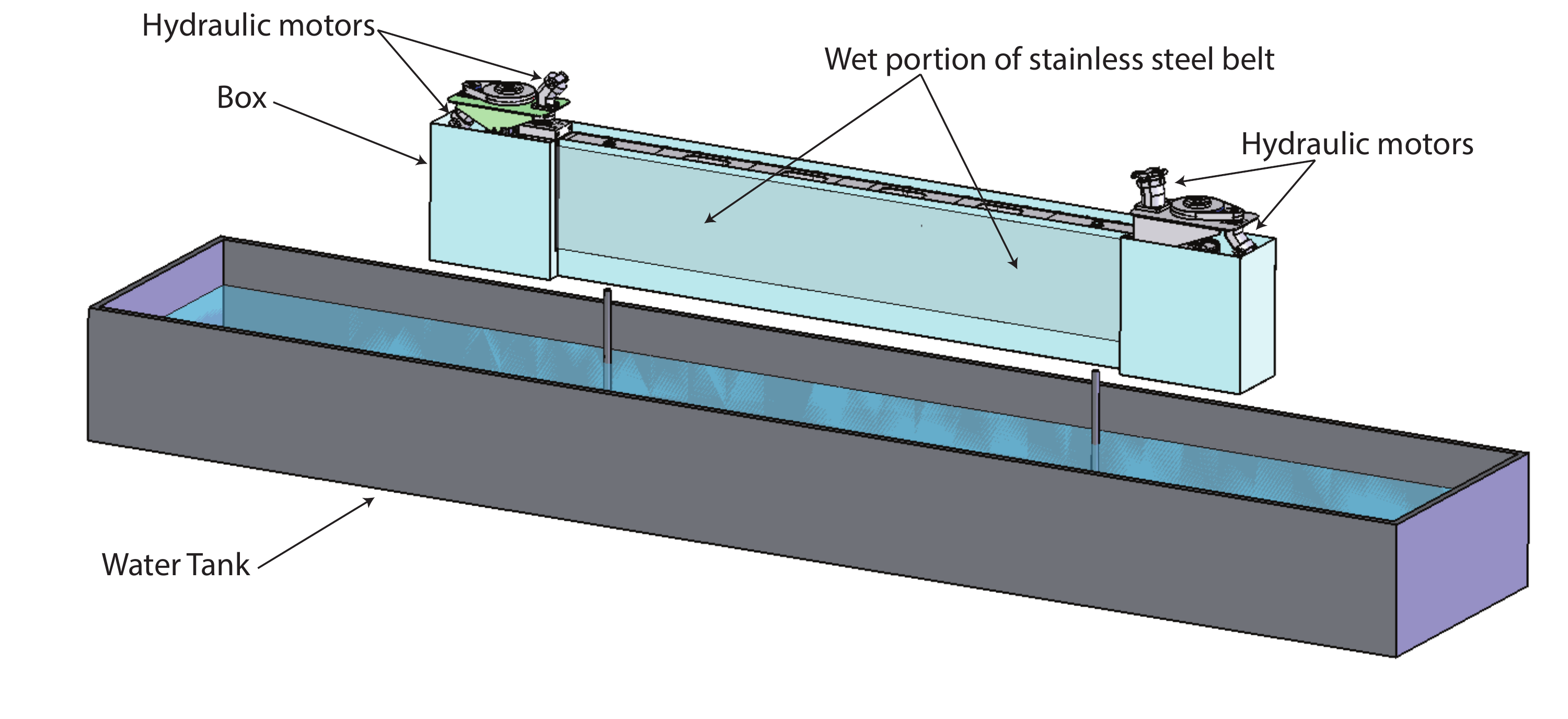}
\end{center}
\vspace*{-0.3in} \caption{Perspective view of the Ship Boundary Layer Simulator (SBLS) and the water tank.} \label{fig:TankSchem}
\end{figure*}

The experiments are performed in an open-surface water tank that is 13.34 m long, 2.37 m wide and 1.32 m deep, see figure~\ref{fig:TankSchem}. The inner surfaces of the tank walls and bottom are composed of 31.8-mm-thick clear Acrylic panels which are supported by a steel frame. The top of the tank is open, offering an unobstructed view of the water surface. Two floor-mounted circular steel pads pierce the bottom of the tank and are used to support the Ship Boundary Layer Simulator (SBLS). The tank includes a water filtration system consisting of a 2.3-m-long skimmer at one end of the tank, a diatomaceous-earth filter and associated pipes and valves. The output line of the skimmer can be directed to the drain or to the filter and from there to return to the tank at the end opposite to the skimmer.  

\begin{figure*}[!htb]
\begin{center}
\includegraphics[trim=0 1in 0in 1.6in,clip=true,width = 6in]{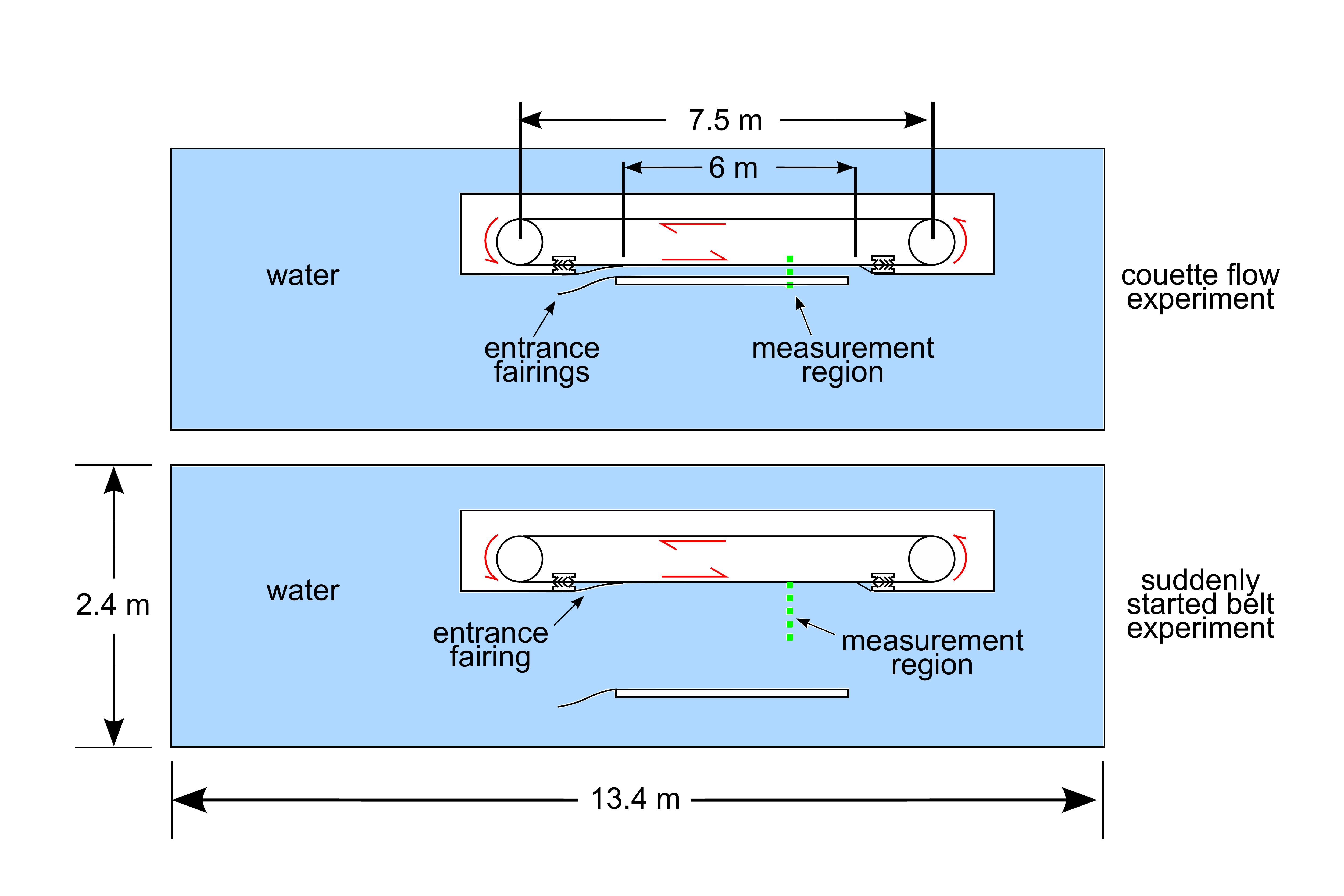}
\end{center}
\vspace*{0in} \caption{Plan view of the SBLS and water tank for (a) the steady state Couette flow experiment and (b) the suddenly started belt experiment.} \label{fig:top_view}
\end{figure*} 

The main functional component of the SBLS is a one-meter-wide 0.8-mm-thick endless stainless steel belt which is driven by two 0.46-meter-diameter, 1.1-meter-long drums whose  rotation axes are vertically oriented and separated by a horizontal distance of approximately 7.5 meters. The drums are each driven by two bent-axis hydraulic motors via toothed belt and pulley systems.  Each drum along with the motors and drive systems form single drive units that are attached to a welded steel frame that maintains the separation between and relative parallel orientation of the drums. The drum drive unit on the left in figure~\ref{fig:TankSchem} is attached to the steel frame via two hydraulic pistons, positioned at the top and bottom of the frame.  The vertical position of the belt  on the drum is controlled actively during each experimental  run by measuring the belt position with a light sensor and tilting the left drum with differential motion of the hydraulic pistons.  Tilting the left drum clockwise (counter clockwise) by small amounts causes the belt move up (down).  During a typical run, the drum position varies by no more than 5~mm.

The assembled SBLS is placed in a stainless steel sheet metal box (called the dry box) as shown in figure~\ref{fig:TankSchem}.  The box keeps the assembly essentially dry, while one of the two roller-to-roller sections of the belt exits the box through a set of seals and travels to the second set of seals near the opposite roller where the belt re-enters the box.  This box was deemed necessary to keep water from corroding the SBLS mechanism and to keep water from being dragged in between the belt and the roller where it might cause the belt to hydroplane off the roller. The lone straight section exposed to water is approximately 6 meters long and pierces the free surface with approximately 0.33 meters of freeboard for the water level used in the present experiments. At the location where the belt leaves the dry box and enters the water, a sheet metal fairing is installed to reduce the flow separation caused by the backwards-facing step associated with the shape of the dry box at this location.

This SBLS device is used in two experiments as depicted in the schematic drawings shown in figure~\ref{fig:top_view}. The first of these experiments is a steady state Couette flow with a free surface. This flow is created by placing a flat, stationary plate parallel to the belt surface with a distance of 4~cm in the gap between the two surfaces.  Like the belt, the top part of the plate pierces the water free surface.   In this experiment, the belt is turned on for a sufficient time for the flow to reach steady state, after which data is collected. The leading edge of the stationary plate includes a fairing to minimize flow entrance effects and the 
measurement location is positioned approximately 100 gap widths downstream of the leading edge of the plate, in order to obtain a  fully developed flow. In these experiments, steady-state belt speeds between 2.8~m/s and 4.0~m/s are investigated.

\begin{figure*}[!ht]
\begin{center}
\begin{tabular}{c}
\includegraphics[trim=0.5in 0.0in 0 0.00in,clip=true,scale=0.6]{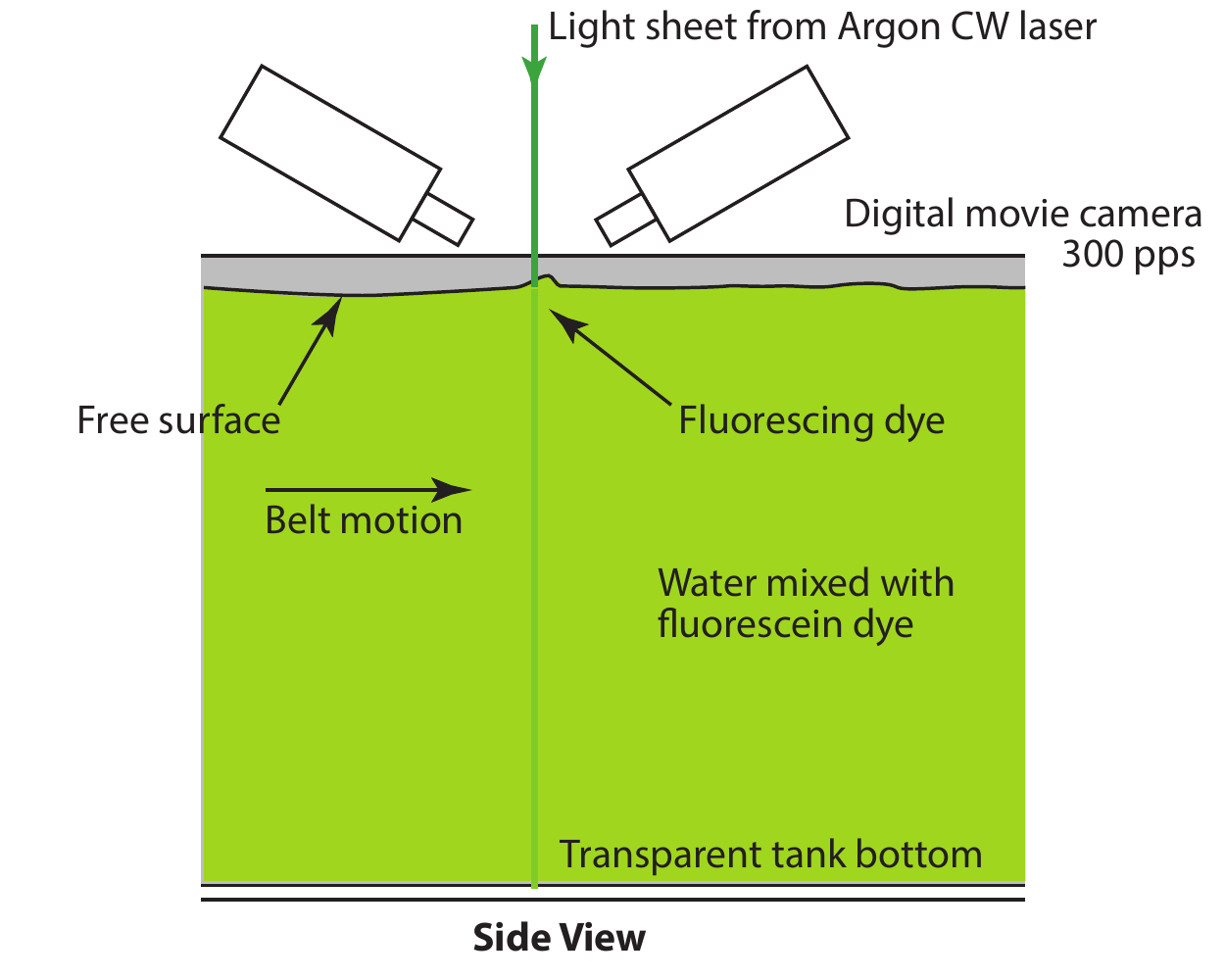}
\includegraphics[trim=0in 0.0in 0 0.00in,clip=true,scale=0.6]{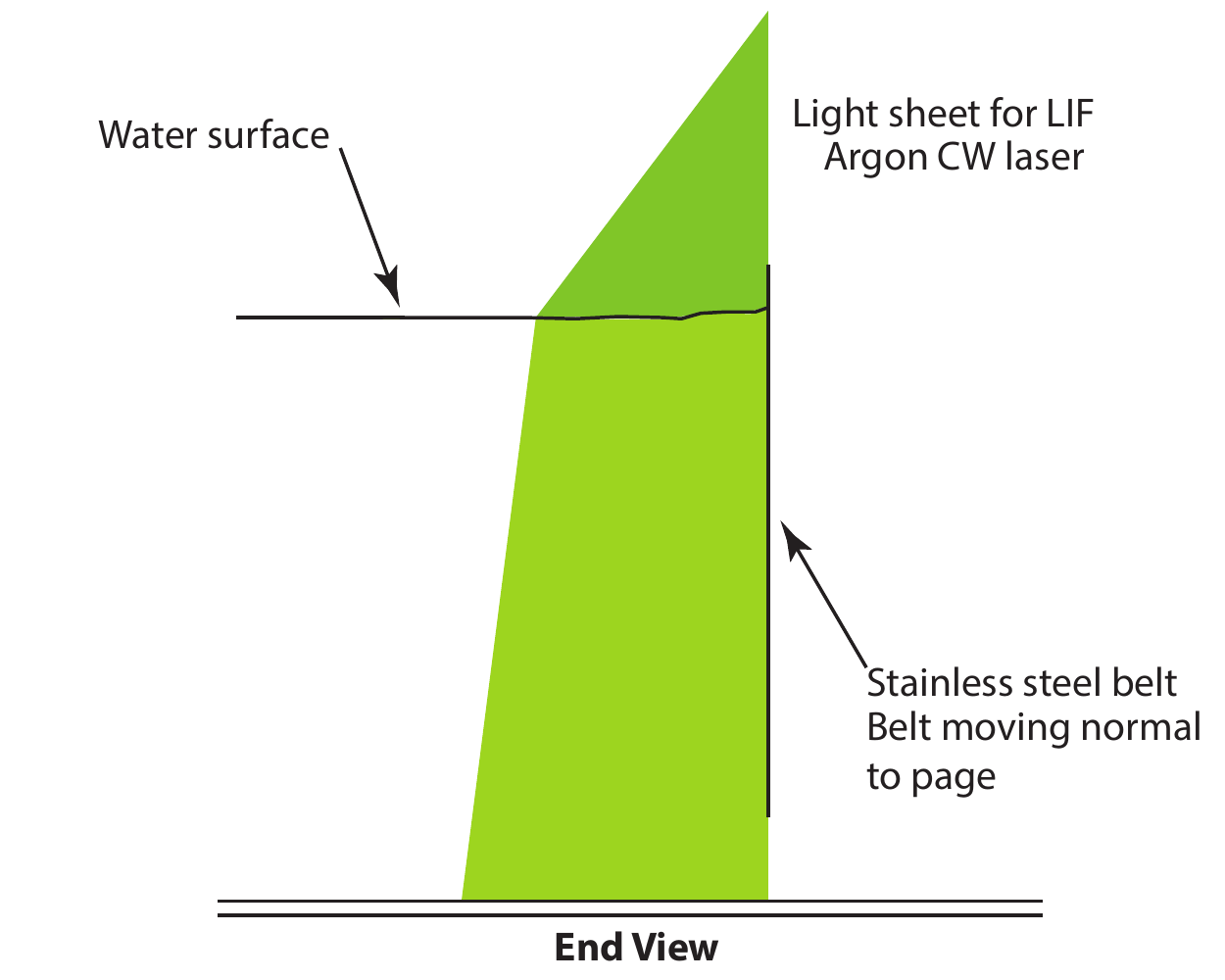}
\end{tabular}
\end{center}
\vspace*{-0.1in} \caption{Schematic drawing showing the set up for the cinematic LIF measurements of the free surface shape.} \label{fig:LIFSchem}
\end{figure*}

In the second set of experiments, the belt is started from rest and the stationary plate is placed 1~m from the belt surface.  
Each run lasts for less than 10~s, and at the end of a run the boundary layer thickness is substantially smaller than the distance from the belt to the fixed plate, thus free field conditions are simulated.  When launched from rest, the belt is capable of acceleration equal to 3 times gravitational acceleration. The time to reach constant belt speed varies depending on the final speed, but is typically less than 0.25 seconds for the experimental conditions used here. Throughout these transient experiments, the belt travel is analogous to the passage of a flat-sided ship that makes no bow waves; the length along the hull is equivalent to the total distance traveled by the belt. Belt speeds ranging from 3 to 5~m/s were used and measurements were continued until a  belt length of 30~m had passed by the measurement site. Therefore, at 3~m/s, experiments run for 10 seconds, while at 5~m/s, experiments run for 6 seconds.

To study the water surface deformation in both experiments, a cinematic Laser Induced Fluorescence (LIF) technique was  utilized, see figure~\ref{fig:LIFSchem}. In this technique, a continuous-wave Argon Ion laser beam is converted to a thin sheet using a system of spherical and cylindrical lenses. This sheet is projected vertically down onto the water surface in an orientation with the plane of the light sheet normal to the plane of the belt.   This laser emits light primarily at wavelengths of 488~nm and 512~nm. The water in the tank is mixed with fluorescein dye at a concentration of about 5~ppm and dye within the light sheet fluoresces. Two cameras view the intersection of the light sheet and the water surface from both upstream and downstream with  viewing angles of approximately 20 degrees from  horizontal. The cameras (Phantom V642 by Vision Research, Inc.) capture 4-Mpixel 12-bit black-and-white images at frame rates up to 1500~Hz.  A long-wavelength-pass optical filter is placed in front of each camera lens.  These filters block out the laser light and transmit the light from the fluorescing dye, thus preventing specular reflections of the laser light from the water surface from entering the camera lenses.  The use of two cameras allows for more accurate surface measurement in the event that the intersection of the light sheet and the water surface is blocked  by large deformations of the free surface between the plane of the light sheet and either camera. The images seen by these cameras shows a sharp line at the intersection of the light sheet with the free surface. Using image processing, instantaneous surface profiles can be extracted from these images.

\begin{figure}[!hb]
\begin{center}
\includegraphics[trim=0.8in 0.3in 0 0.00in,clip=true,scale=0.6]{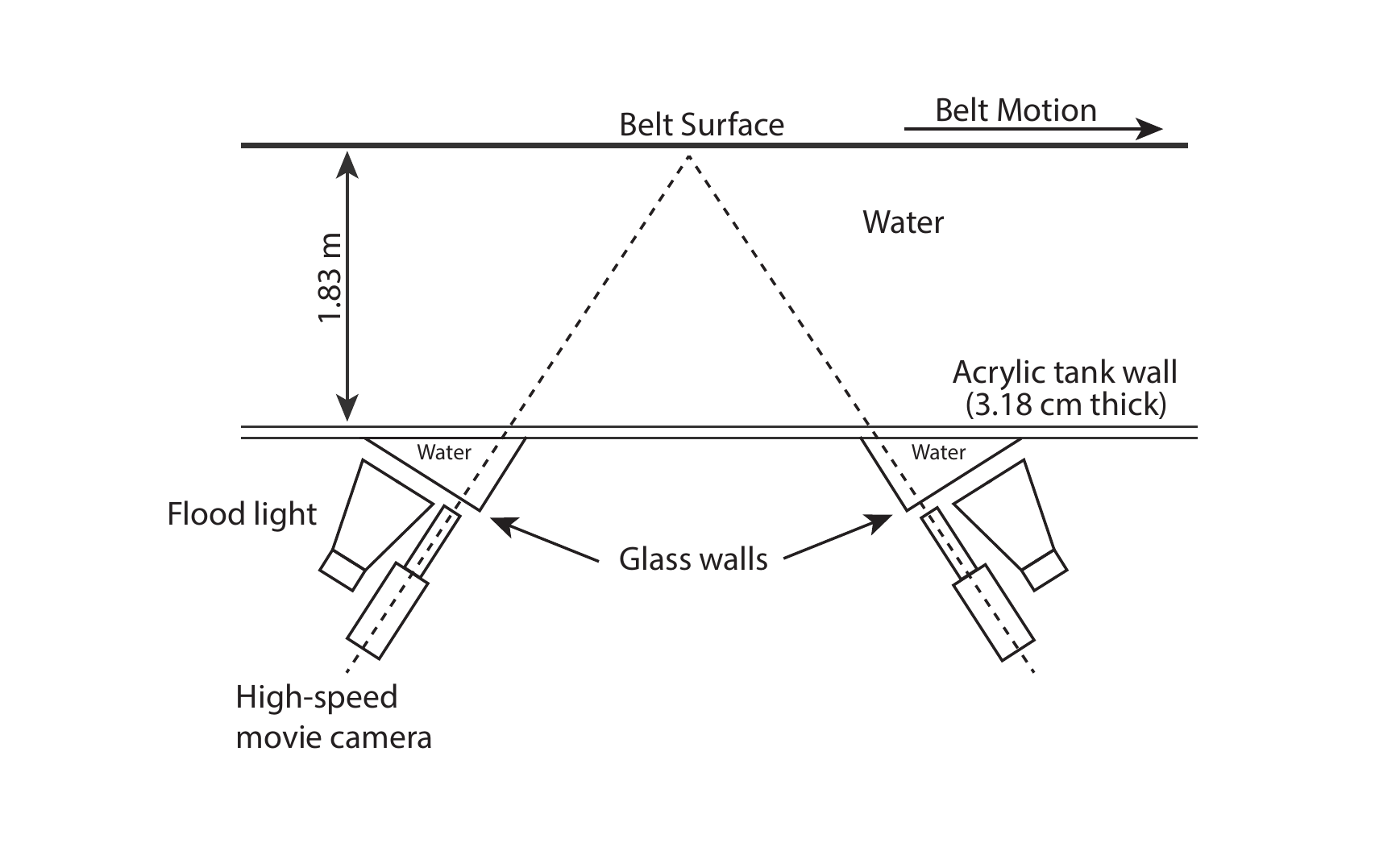}
\end{center}
\vspace*{-0.3in} \caption{A schematic of the underwater stereo bubble measurement system} \label{fig:BubbleSchem}
\end{figure}

In addition to the above-described surface profile measurements, observation of the entrainment and motions of air bubbles were undertaken with a cinematic stereo camera system in the experiments with the belt starting from rest, see schematic in figure~\ref{fig:BubbleSchem}.   In this system, the cameras view the flow just under the water free surface adjacent to the belt.  The cameras are set up along the side wall of the tank that is parallel to the belt with horizontal viewing directions and lines of sight rotated $\pm 30^\circ$  about a vertical plane that is normal to the belt surface.   A photoflood light with a translucent screen is placed next to each camera.  Since the belt surface is moderately reflective, each light provides the illumination for the opposite camera.  The cameras (lights) view (illuminate) the scene through water-filled prism boxes that are set to create a straight line of sight that is perpendicular to the side wall of the prism tank.      Each camera is also equipped with a Scheimpflug lens mount, which inclines the sensor plane relative to the lens plane so that each camera may focus on a plane parallel to the belt despite viewing from a displaced angle. Using two cameras at offset angles allows for a more accurate characterization of the non-axisymmetric free surface features found during entrainment events. Additionally, utilizing the principles of parallax, three dimensional bubble positions can be determined.  Though quantitative measurements are planned using this system in the near future, in the present paper, only qualitative observations will be described.

\section{3. Results and Discussion}

In this section, the  results for both the Couette flow and the suddenly started belt experiments are presented and discussed. The  surface profile measurements from the Couette flow experiments are addressed first and then followed by a description of the surface profile measurements and air entrainment observations in  the suddenly started belt experiments.  Finally, a discussion of air-entrainment boundaries for both experiments is given.

\subsection{Couette Flow Experiment}
\subsubsection{Surface Profiles}

\begin{figure}[!htb]
\begin{center}
\begin{tabular}{c}
(a)\\
\includegraphics[trim=0 0.0in 0 0.00in,clip=true,scale=0.22]{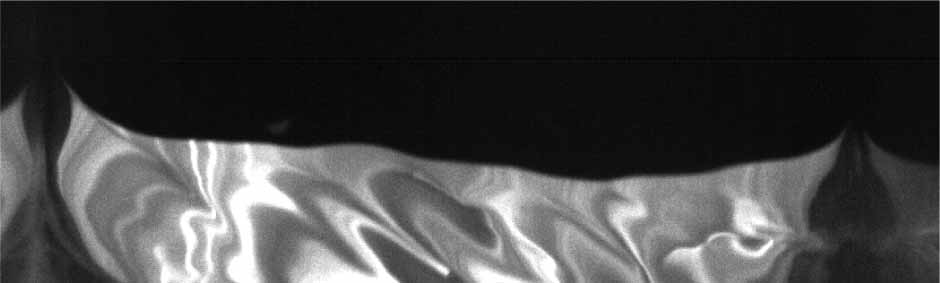}\\
(b)\\
\includegraphics[trim=0 0.0in 0 0.00in,clip=true,scale=0.22]{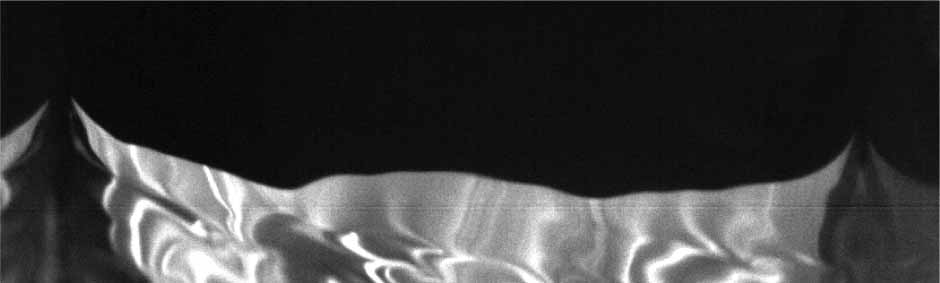}\\
(c)\\
\includegraphics[trim=0 0.0in 0 0.00in,clip=true,scale=0.22]{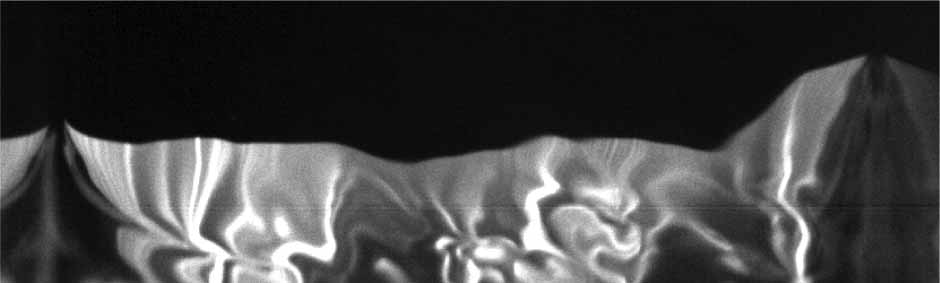}\\
(d)\\
\includegraphics[trim=0 0.0in 0 0.00in,clip=true,scale=0.22]{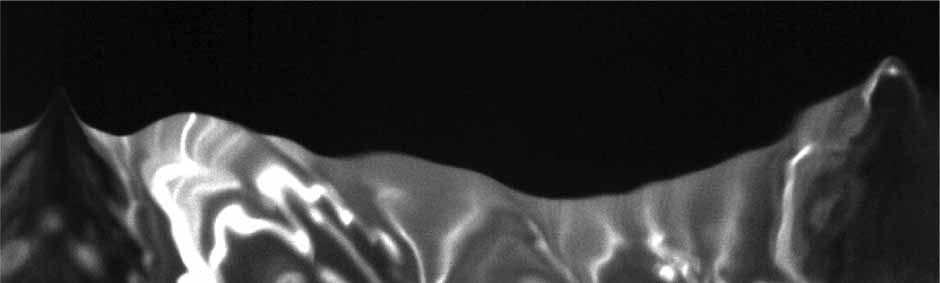}\\
\end{tabular}
\end{center}
\vspace*{-0.2in} \caption{Four images from separate Couette flow experiments, depicting steady state belt speeds of (a) 2.8~m/s (b) 3.2~m/s (c) 3.6~m/s and (d) 4.0~m/s. The horizontal field of view for each image is approximately 5~cm.}\label{fig:couette_snapshot}
\end{figure}

In this set of experiments, the belt speed is varied between 2.8~m/s to 4.0~m/s which corresponds to a Froud number range of  3.01 to 4.31 based on the definition $Fr=U/(2gH)^{1/2}$ where U is the belt speed, g the gravitational acceleration and H the gap width. Single LIF images taken from the high-speed movies for belt speeds of 
2.8, 3.2, 3.6 and 4.0~m/s are shown in figures~\ref{fig:couette_snapshot}(a), (b), (c) and (d), respectively. These images were taken with the upstream camera;  the belt is on the left side of the image and the fixed plate is on the right.  The belt is moving away from the camera, i.e., into the page in the images.    The light sheet is projected  from above and the glowing dye at the intersection of the light sheet and the water surface is seen as a high-contrast boundary between the upper dark region and the lower light region in each image.  The shape of this boundary is measured quantitatively using image processing techniques. (The automated version of these techniques worked reliably except for regions within about 4~mm of the belt or wall.  In these regions, the profiles were determined by eye.)  The bright area below the boundary is created by the glowing fluorescent in the underwater portion of the light sheet.  The complex light intensity pattern here is created by a combination of the refraction of the laser light sheet as it passes down through the water surface and the refraction of  the light from the glowing underwater dye as the light passes up through the water surface between the light sheet and the camera, on its way to the lens.   This part of the image cannot be quantitatively analyzed.  As can be seen from the images, the surface height fluctuations increase steadily with belt speed,  but wave breaking and air entrainment are not evident, even at the highest speed.

\begin{figure*}[!htb]
\begin{center}
\begin{tabular}{ccc}
(a)&(b)&(c)\\
\includegraphics[trim=0.4in 0.0in 0.65in 0.5in,clip=true,scale=0.32]{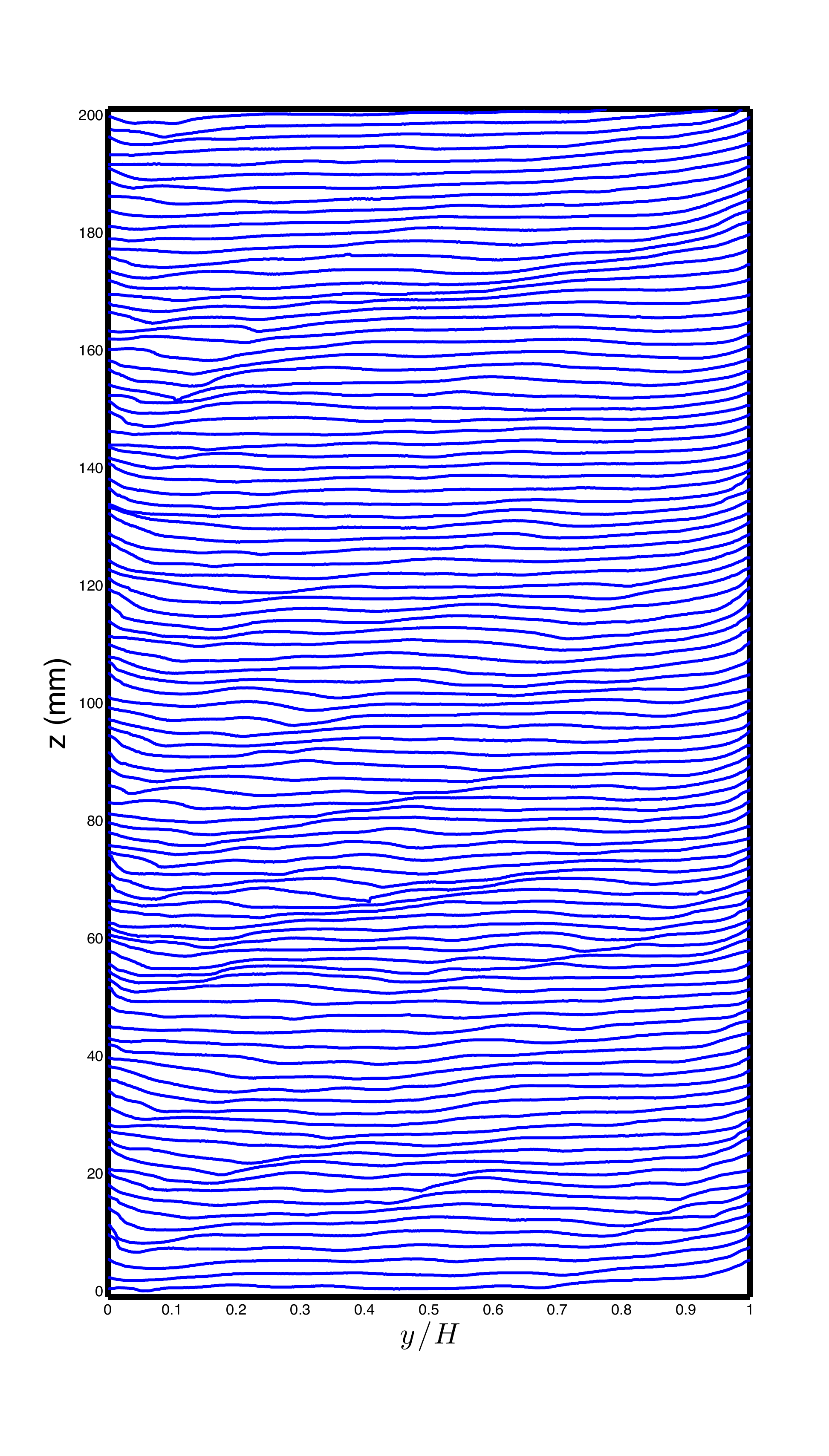}&
\includegraphics[trim=0.98in 0.0in 0.65in 0.5in,clip=true,scale=0.32]{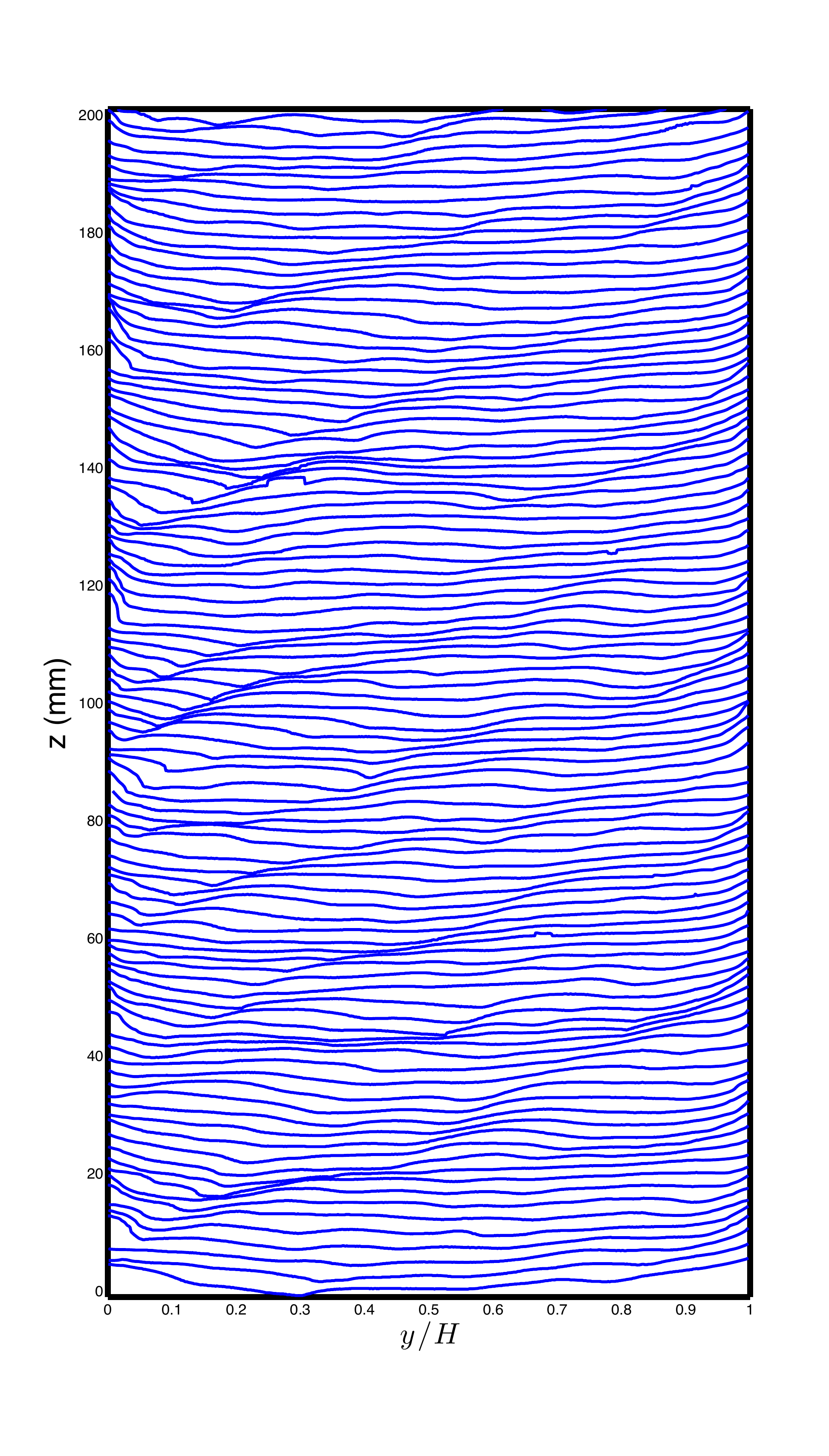}&
\includegraphics[trim=0.98in 0.0in 0.65in 0.5in,clip=true,scale=0.32]{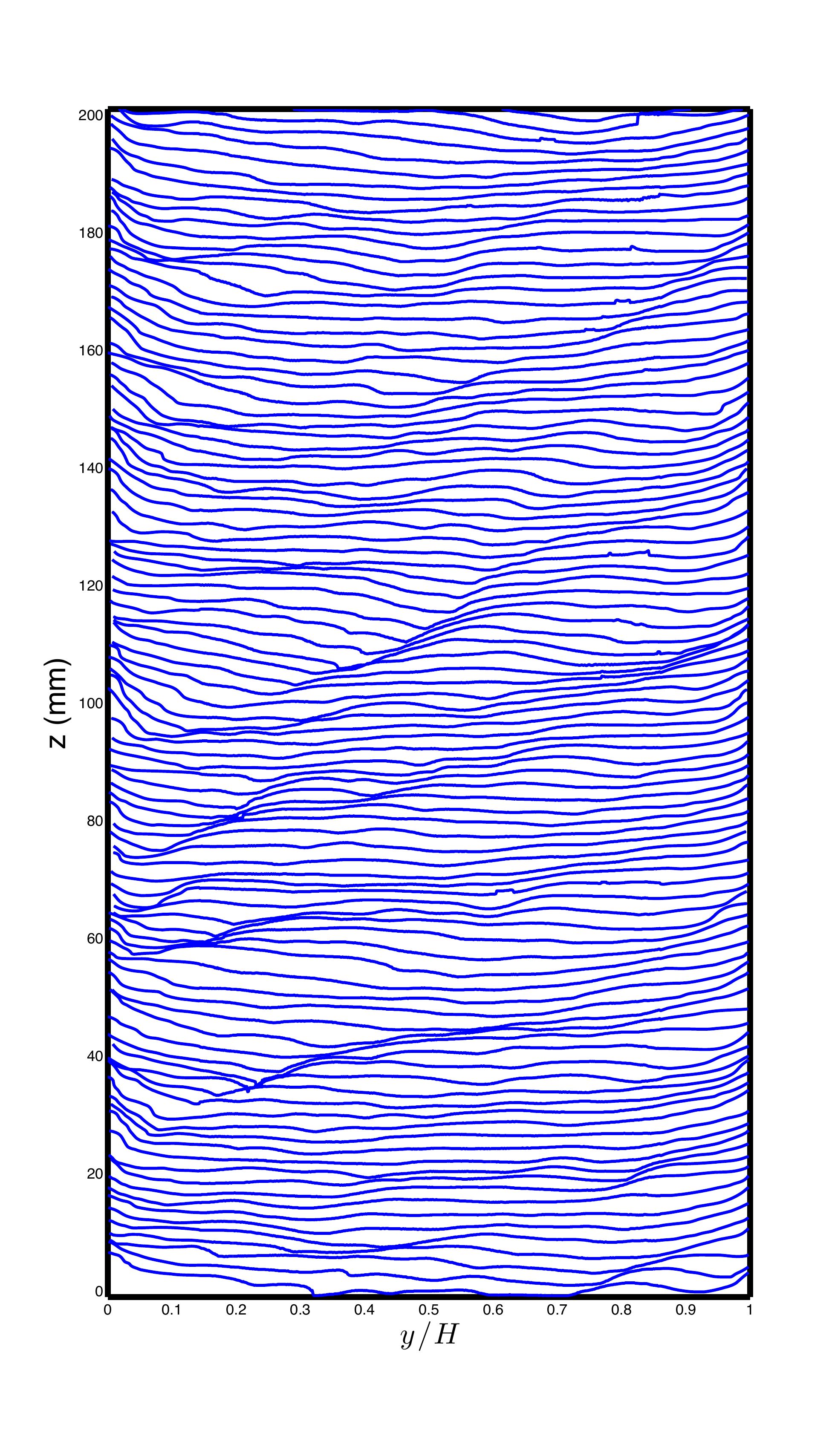}\\
\end{tabular}
\end{center}
\vspace*{-0.2in} \caption{A series of profile history plots, with each depicting a surface profile history in time throughout the Couette flow experiment for (a) 3.0~m/s, (b) 3.4~m/s, and (c) 3.8~m/s. In each plot, profiles are offset vertically by 2~mm from the previous profile so that the evolution of features can be seen.}\label{fig:couette_profiles}
\end{figure*}

Instantaneous surface profiles were extracted from  5,000 images (corresponding to 10~s in real time) from one of the  high-speed movies at each belt speed.   A series of about 100 surface profiles for each of three belt speeds is shown in figure~\ref{fig:couette_profiles}. In the plots, each profile is shifted up by 2~mm from the previous profile to reduce overlap and so that the spatio-temporal evolution of surface features can be shown. The horizontal axis, $y/H$, is the distance from the belt surface normalized by the gap width, $H$; the fixed plate is located at $y/H=1.0$. If one were to draw lines connecting surface profile features like the local crests from one profile to the next, the slopes of these lines would correspond to the horizontal speed of these features in a direction normal to the belt. It can be seen from the plots that the visually dominant  features travel from the belt side to the plate side and the surface height fluctuations become bigger as the belt speed is increased.

\subsubsection{Surface Fluctuations}

Surface profiles extracted from the images were used to analyze the temporal and spatial characteristics of the surface height fluctuations. Though each high-speed LIF movie from which the profiles were extracted was taken   long after the belt  started from rest and had reached a steady state speed, there were some low-frequency overall oscillations  in the surface profile records.   The cause of these oscillations is at present unknown, but is being explored in ongoing experiments.  Since the surface ripples have a relatively high frequency, it was decided to remove the low-frequency oscillation from the data records by high-pass filtering before further processing.   Thus, for each pixel column across the Couette flow gap in the images, a fast Fourier transform of a sequence of 5000 data points was obtained. It was observed that for all cases the background oscillations had a frequency of about 0.4~Hz, thus the cut off for the high-pass filter was taken as 0.5~Hz.

\begin{figure*}[!ht]
\begin{center}
\begin{tabular}{ccc}
(a)&(b)\\
\includegraphics[trim=0.4in 0.0in 0.5in 0in,clip=true,scale=0.4]{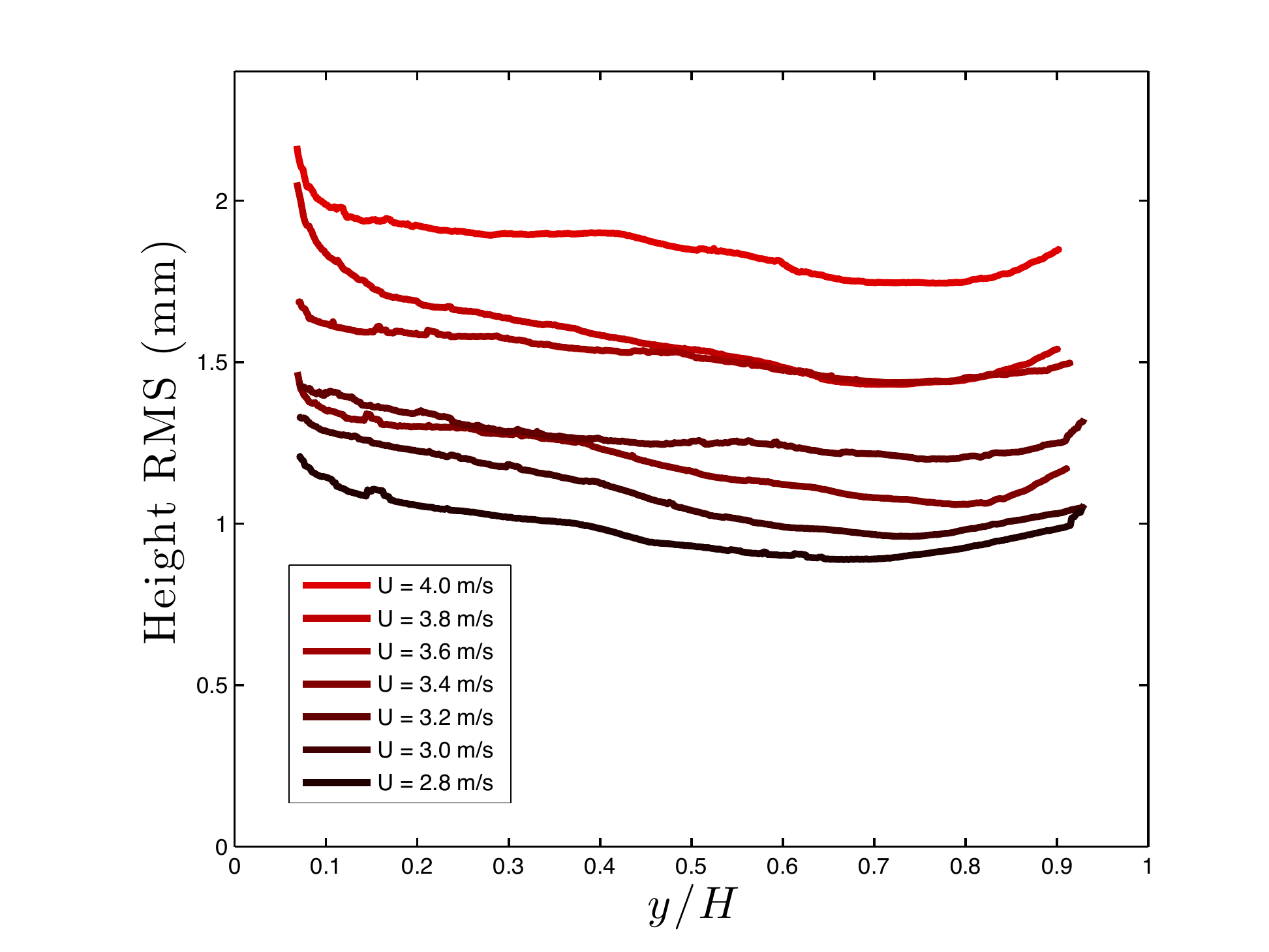}&
\includegraphics[trim=0.4in 0.0in 0.5in 0in,clip=true,scale=0.4]{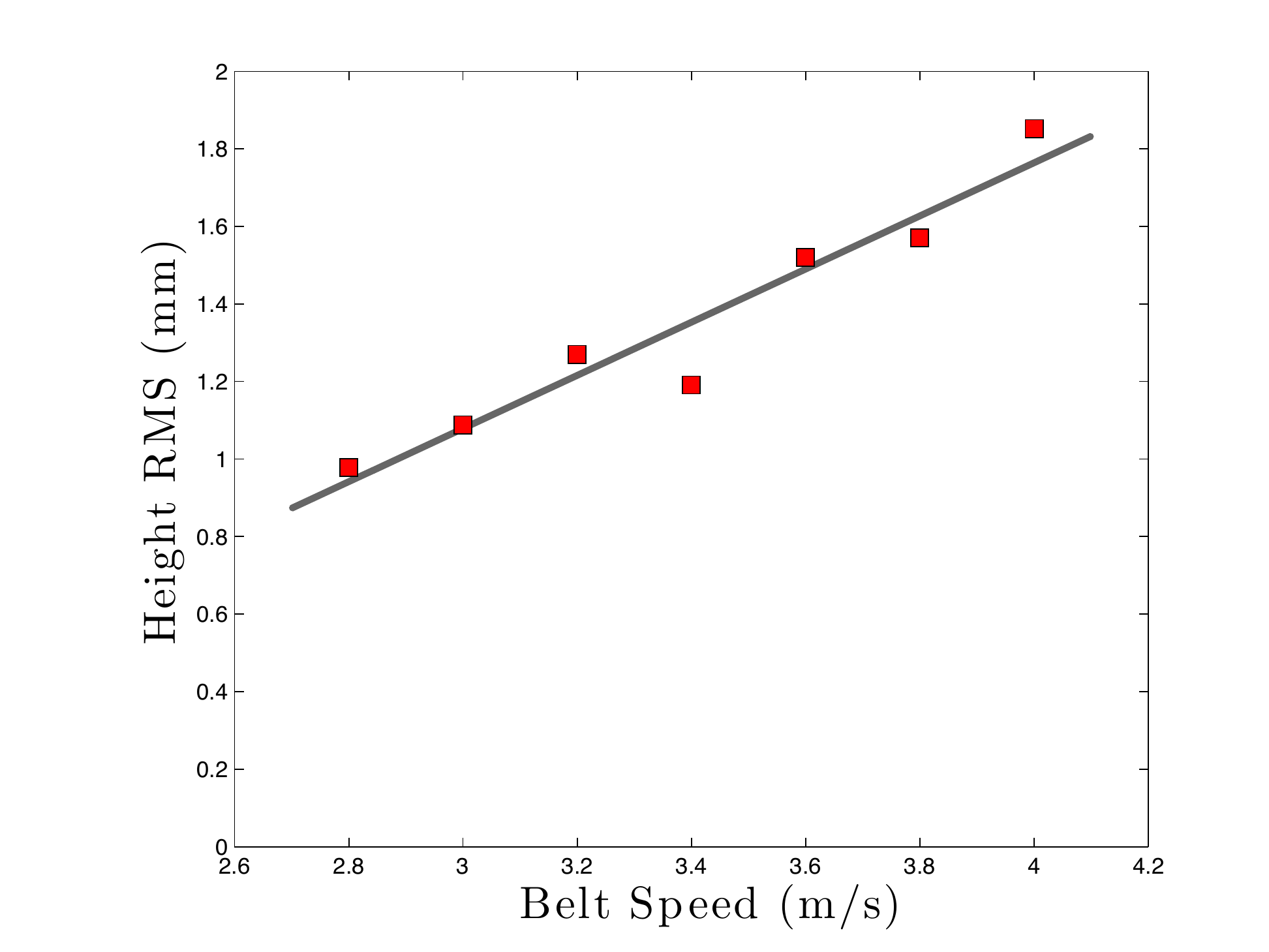}&
\end{tabular}
\end{center}
\vspace*{-0.3in} \caption{(a) Surface height RMS values depicted across the gap for a range of belt speeds. Each profile is averaged across 5000 frames. (b) Total surface height RMS averaged over the gap width for a range of belt speeds.} \label{fig:couette_rms_vs_y}
\end{figure*}

Curves of the temporal average  of the RMS of the filtered surface height fluctuations  versus normalized position, $y/H$, from the belt surface  for each of the belt speeds used in this study are shown in figure~\ref{fig:couette_rms_vs_y} (a). The moving wall is located at position $y/H = 0$ and the stationary plate is at $y/H  = 1$.   In the regions next to the belt and the fixed wall, the water surface profile data has not yet been extracted for all of these belt speeds and so is not plotted for any cases.  It is observed that the RMS height fluctuation is higher on the belt side of the gap, but it increases locally at both walls. The average over the gap width of the RMS surface data plotted in figure~\ref{fig:couette_rms_vs_y} (a) is plotted versus belt speed in figure~\ref{fig:couette_rms_vs_y} (b).  In general, the RMS value increases with the belt speed as expected.  It is also found that with the current amount of processed data, the curves are still a bit noisy.  Additional measurements and data processing are planned for the near future in order to  eliminate this problem.

\begin{figure*}[!htb]
\begin{center}
\begin{tabular}{ccc}
(a)&(b)\\
\includegraphics[trim=0.4in 0.0in 0.5in 0in,clip=true,scale=0.4]{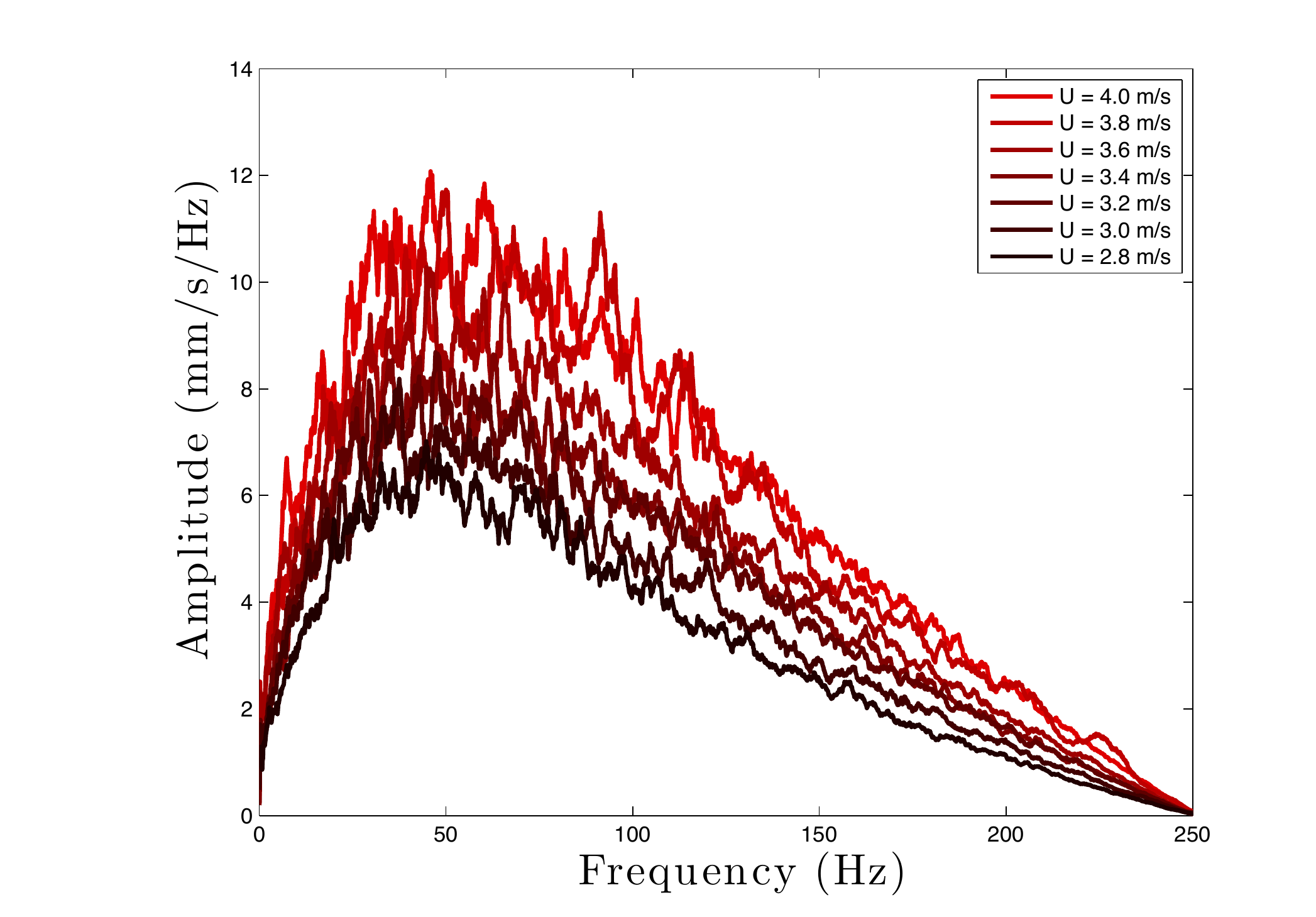}&
\includegraphics[trim=0.4in 0.0in 0.5in 0in,clip=true,scale=0.4]{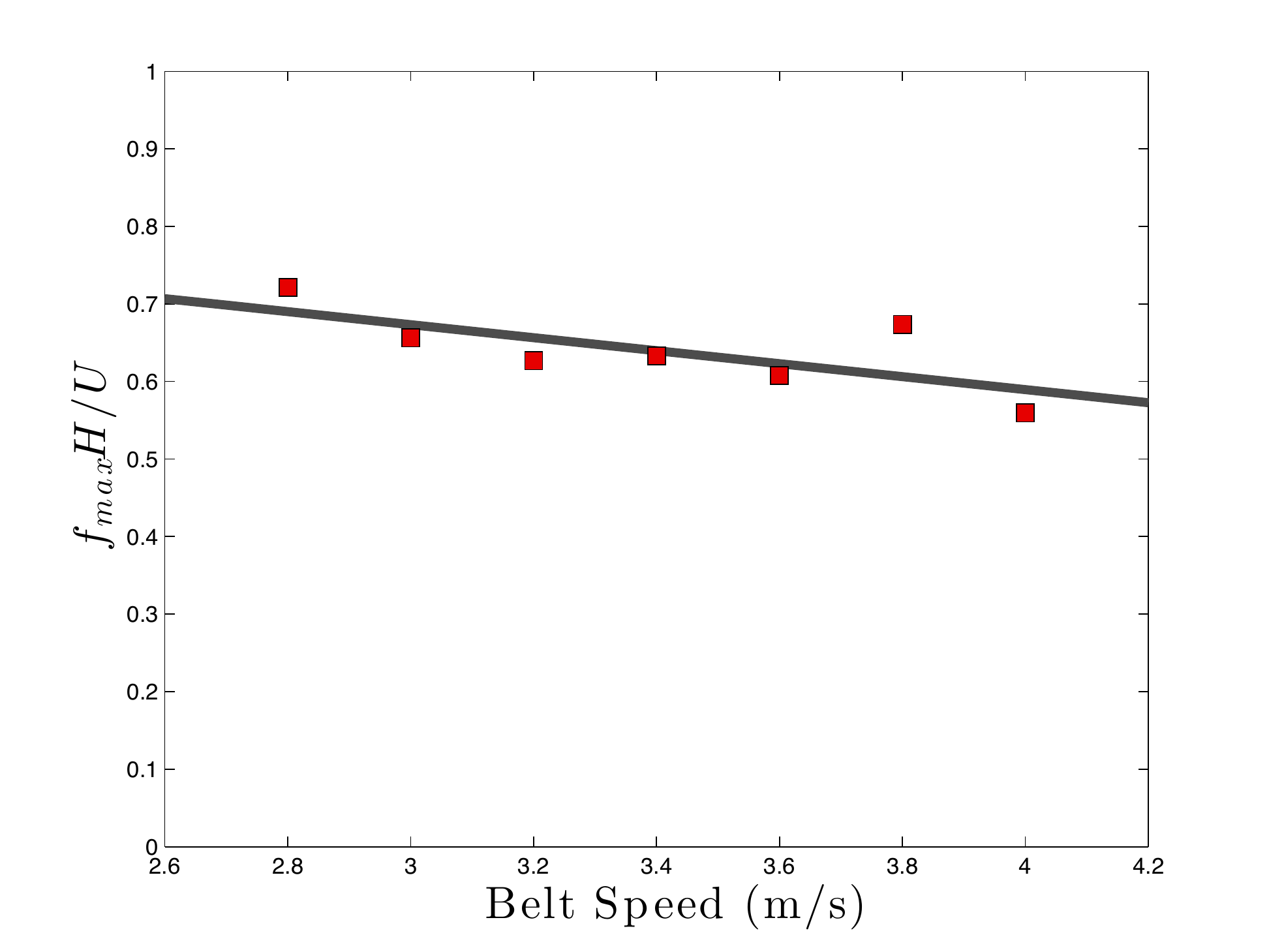}&
\end{tabular}
\end{center}
\vspace*{-0.3in} \caption{(a) Fourier spectrum of time derivative of height fluctuations averaged over the gap width. (b) Frequency of the maximum amplitude versus the belt speed} \label{fig:couette_v_fft_1}
\end{figure*} 

\begin{figure*}[!htb]
\begin{center}
\begin{tabular}{ccc}
(a)&(b)\\
\includegraphics[trim=0.4in 0.0in 0.5in 0in,clip=true,scale=0.4]{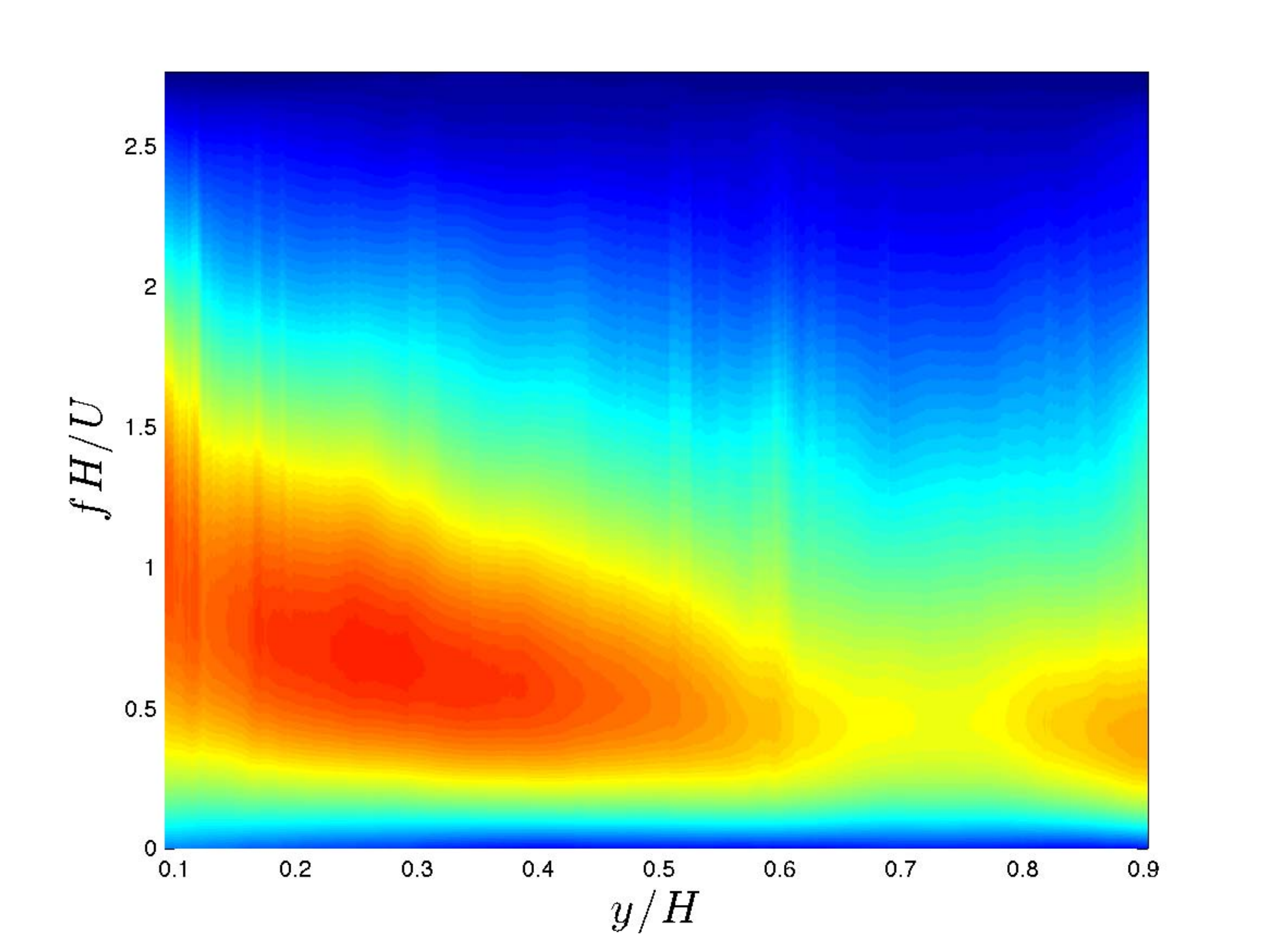}&
\includegraphics[trim=0.4in 0.0in 0.5in 0in,clip=true,scale=0.4]{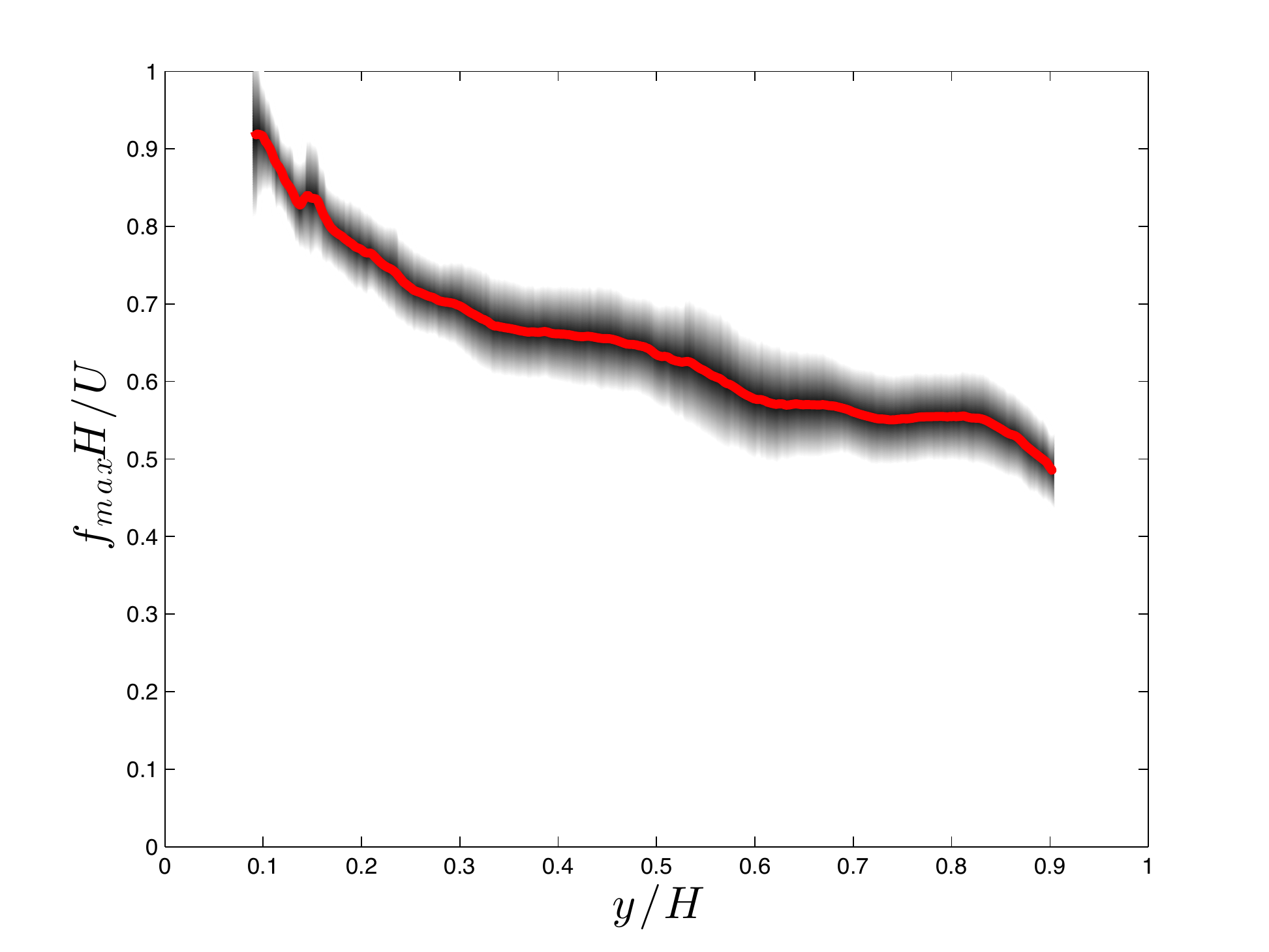}&
\end{tabular}
\end{center}
\vspace*{-0.3in} \caption{(a) Fourier spectrum of time derivative of height fluctuations averaged over the gap width for $U=4.0$~m/s. (b) Non-dimensional frequency of the maximum amplitude ($f_{max}H/U$) versus non-dimensional position ($y/H$) across the gap and averaged over the seven belt speeds.} \label{fig:couette_v_fft_2}
\end{figure*} 


The spectrum of the temporal derivative of the surface height fluctuations ($\partial\eta/\partial t$) was computed at each pixel location across the gap width and then averaged over the gap width.  Plots of these averaged spectra for all seven speeds are given in figure~\ref{fig:couette_v_fft_1}(a).  Though the spectral noise level is a bit high with the amount of data processed so far, it can be seen than the spectral level generally increases with belt speed and that the spectral amplitude has a peak in the range of frequencies from 40 to 50~Hz.  In order to explore how this frequency varies with belt speed, a 6th order polynomial was fitted to each spectrum and the dimensionless frequency ($f_{max}H/U$) of the peak of each polynomial is plotted versus belt speed in figure~\ref{fig:couette_v_fft_1}(b).  The values of $f_{max}H/U$ vary from about 0.7 at the lowest belt speed to about 0.6 at the highest belt speed.

The 6th order polynomial fit to the spectrum for $U=4.0$~m/s as a function of position across the channel width is shown in figure~\ref{fig:couette_v_fft_2}(a).  In this plot, the vertical axis is non-dimensional  frequency ($fH/U$), the horizontal axis in non-dimensional position across the gap width, $y/H$, and the spectral intensity is given by the color according to the scale on the right.  The spectrum varies across the gap width with non-dimensional frequency of the spectral peak  decreasing from about 0.9 near the belt to about 0.5 near the fixed wall.  For reference, it should be kept in mind that a  dimensionless frequency of about 1 is consistent with the model of  surface disturbances of stream wise length scale $H$ being convected past the measure station at the belt speed, $U$. The peak spectral intensity is highest at the belt surface  and reaches a minimum at about $y/H=0.7$.  The reason for this lack of symmetry of the spectra across the channel width, including the possibility that at the measurement location along the channel length (about $x/H = 100$) the flow is not yet fully developed, is presently being explored with additional  experiments.  A plot of the non-dimensional frequency ($f_{max}H/U$) of the maximum of the spectral intensity versus position across the gap and averaged over all belt speeds is given in figure~\ref{fig:couette_v_fft_2}(b).  In this plot, the red curve is the average and the grey area above and below the curve is the standard deviation over the belt speeds.  As can be seen from the plot, the non-dimensional frequency varies from about 0.9 near the belt surface to about 0.5 near the fixed wall. While the level of noise precludes a discussion of the exact shape of this curve, the asymmetry across the gap is evident.

\subsection{Suddenly Started Belt Experiment}

\begin{figure*}[!htb]
\begin{center}
\begin{tabular}{c}
(a)\\
\includegraphics[trim=0 0.0in 0 0.00in,clip=true,scale=0.15]{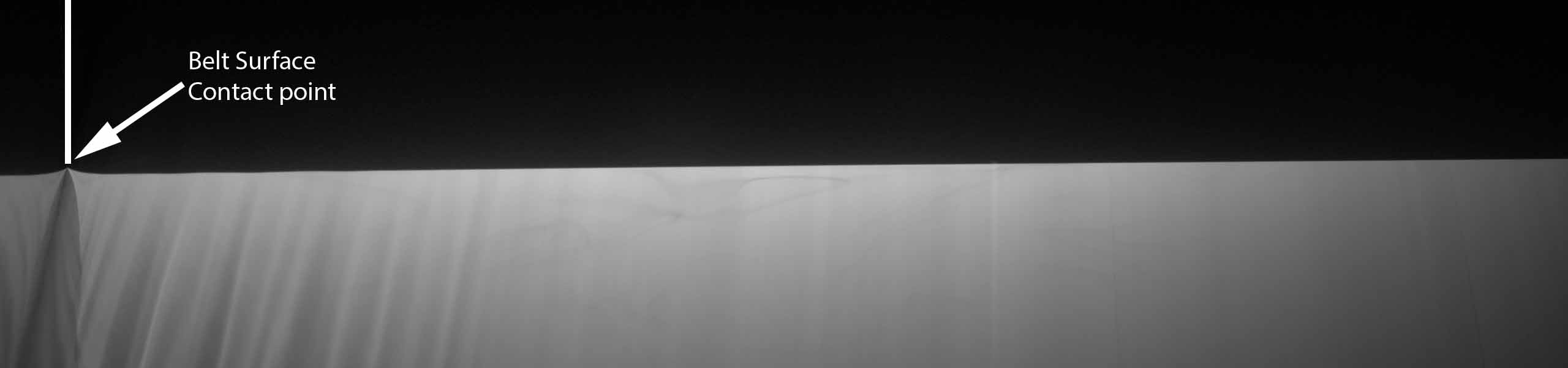}\\
(b)\\
\includegraphics[trim=0 0.0in 0 0.00in,clip=true,scale=0.15]{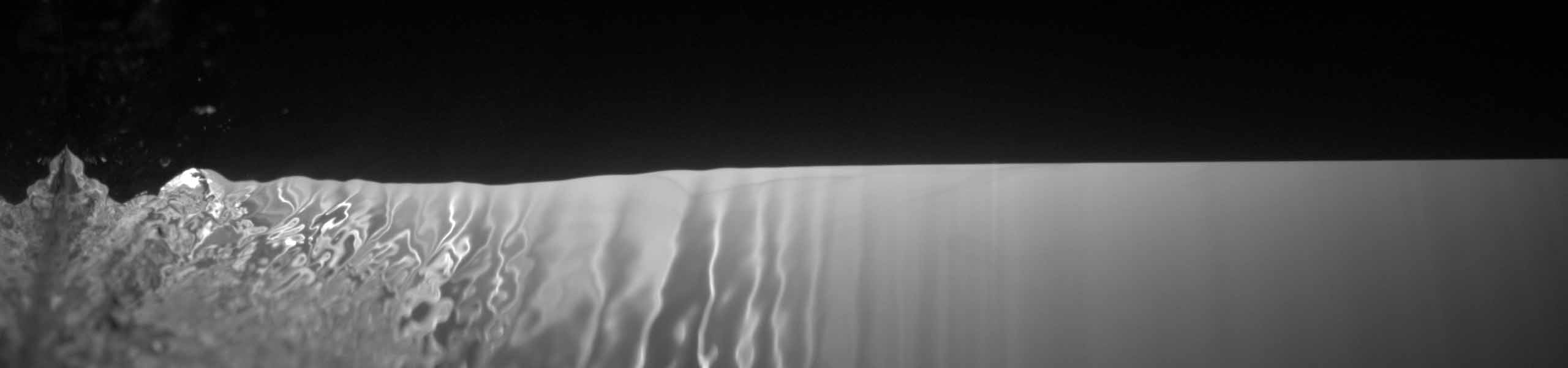}\\
(c)\\
\includegraphics[trim=0 0.0in 0 0.00in,clip=true,scale=0.15]{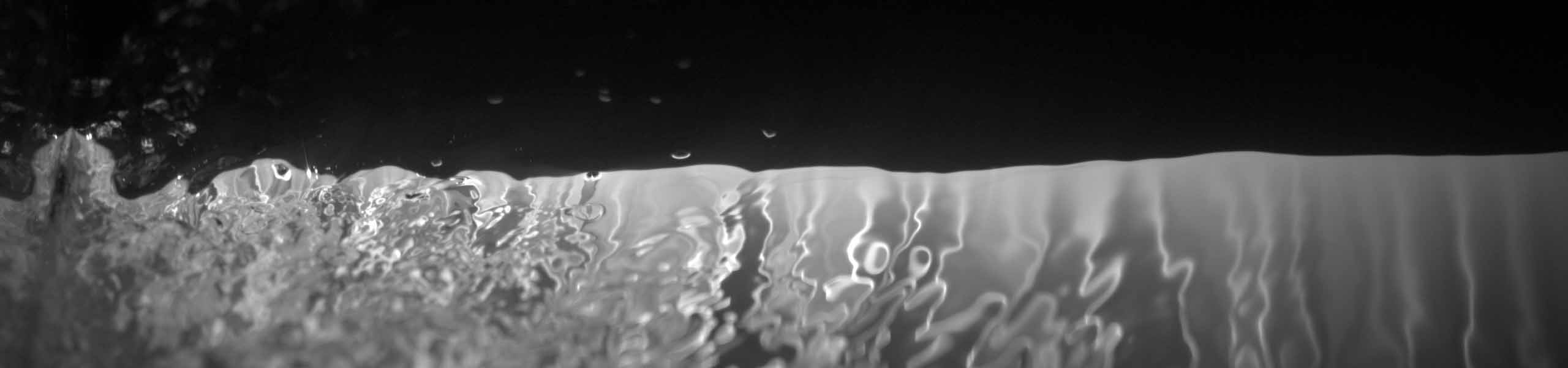}\\
(d)\\
\includegraphics[trim=0 0.0in 0 0.00in,clip=true,scale=0.15]{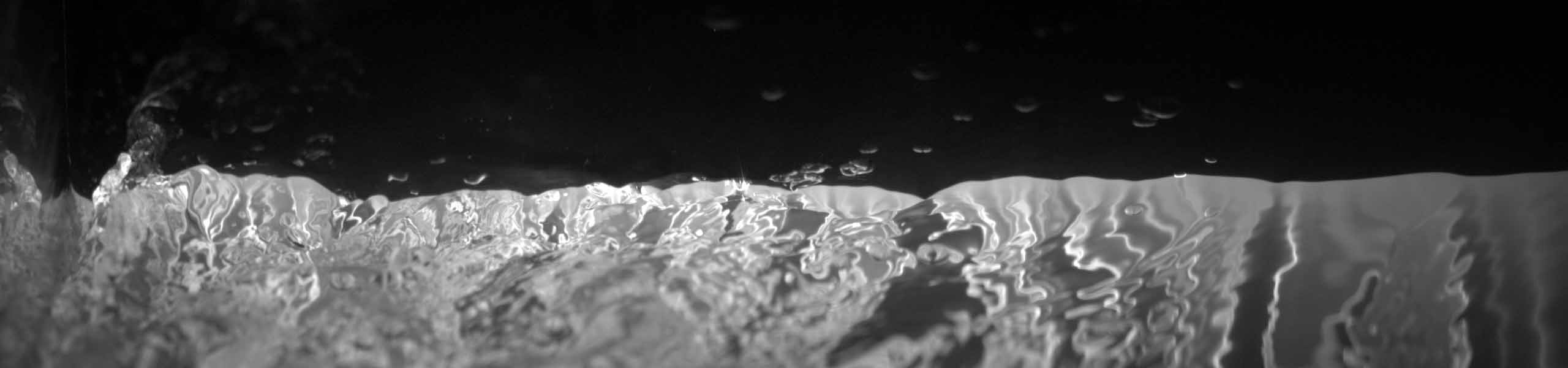}\\
(e)\\
\includegraphics[trim=0 0.0in 0 0.00in,clip=true,scale=0.15]{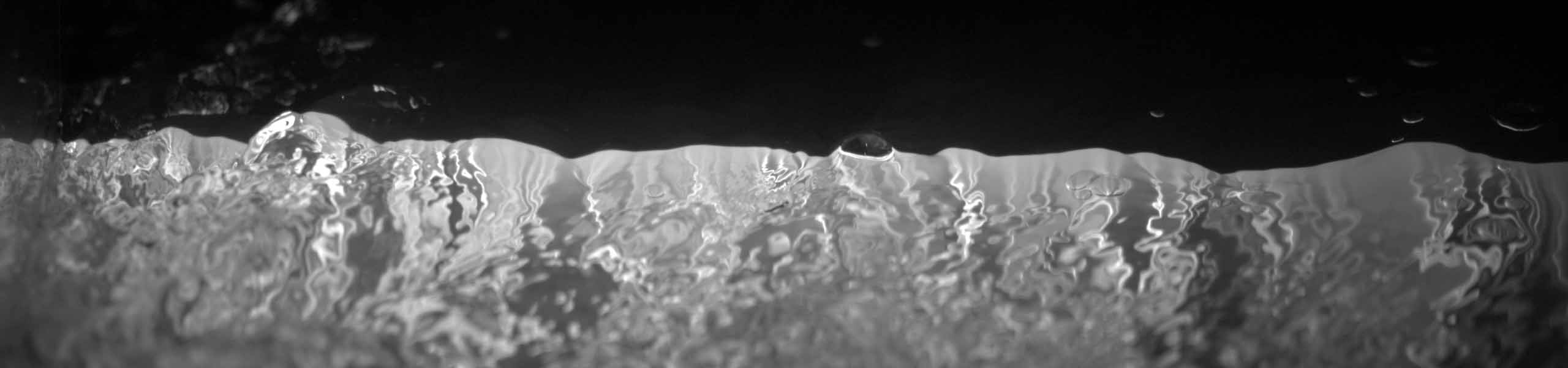}\\
\end{tabular}
\end{center}
\vspace*{-0.2in} \caption{A sequence of five images from a high speed movie of the free surface during a belt launch to 5~m/s. These images are taken at equivalent belt lengths of (a) 0~m (b) 5~m (c) 10~m (d) 15~m and (e) 20~m from the bow of the ship. The high reflectivity of the stainless steel belt makes it appear as a symmetry plane on the left side of the images. The horizontal field of view for these images is approximately 31~cm.}\label{fig:overall}
\end{figure*}

\begin{figure*}[!htb]
\begin{center}
\includegraphics[trim=0 0in 0 0.5in, clip=true,scale=0.6]{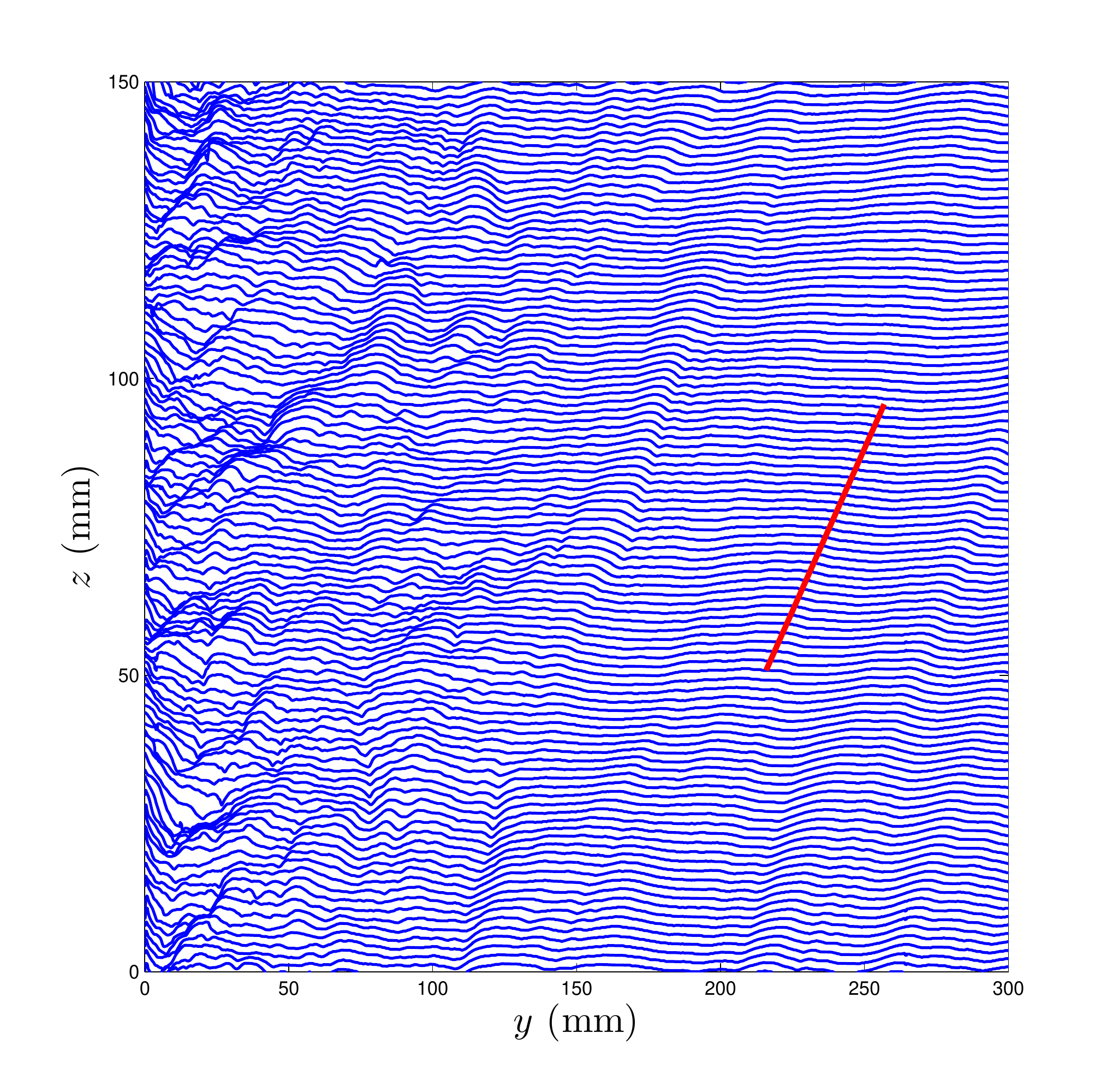}\\
\end{center}
\vspace*{-0.4in} \caption{Sequence of profiles of the water surface during belt launch to 3~m/s.  The time between profiles is 4~ms and each profile is shifted up 2~mm from the previous profile to reduce overlap and show propagation of surface features throughout time. The belt is positioned at the left edge of the image ($y =0$). The red line drawn over the image gives an estimation of the wall-normal propagation speed of surface features. In this case, the red line corresponds to a wall-normal velocity of approximately 34~cm/s.} \label{fig:profiles}
\end{figure*} 

\subsubsection{Surface Profiles}
Five LIF images of the water free surface next to the belt for an experimental run with $U=5$~m/s are shown in figure~\ref{fig:overall}.  Though the camera recorded 1,000 frames per second, the five images in the figure are spaced by an interval of 1~s with the first image (a) taken at 0.0~s.  Here and in the following, rather than refer to images and data by the time after the belt started moving, we refer to them by the distance, $x=Ut$, from the leading edge of an equivalent flat plate moving at constant speed $U$.  Thus, the images in figure~\ref{fig:overall}(a), (b), (c), (d) and (e) correspond to $x=$ 0.0, 5.0, 10.0, 15.0 and 20.0~m, respectively.   As discussed in the experimental details section, the plane of the vertical light sheet is oriented normal to the belt surface and the cameras look parallel to the belt surface and down at the water surface at a small angle from horizontal. The images in figure~\ref{fig:overall} are from the upstream camera so the belt is near the left side of each image and is moving into the page. 
The sharp boundary between the upper dark and lower bright regions of the images is the intersection of the light sheet and the water surface.   The upper regions of the later  images contain light scattered from roughness features on the water surface behind the light sheet.  These roughness features include bubbles that  appear to be floating on the water surface and moving primarily in the direction of the belt motion. 
The position of the belt is marked on the left side of image (a) and the intensity pattern to the left of this location, is a reflection of the light pattern on the right, in the smooth surface of the belt.  
This line of symmetry gives a good indication of the position of the belt in each image.   The the high-contrast boundary  where the light sheet  intersects the free surface is analyzed quantitatively to extract instantaneous surface shapes. 

In the experimental run corresponding to the images in figures~\ref{fig:overall}, the belt launches from rest and reaches a steady state velocity of 5~m/s, after traveling for only about 0.2~s, equivalent to $x= 1.0$~m. It can be seen from these images that surface height fluctuations (ripples)  are created close to the belt surface, at the left side of each image, and propagate away from the belt (to the right). As time passes, the surface height fluctuations grow dramatically and eventually surface breaking and air entrainment events begin to occur, resulting in bubble and droplet production.  These events are more easily observed in high-speed LIF movies.

An example series of water surface profiles from a run with $U=3.0$~m/s over a range of $x$ from  21~m to 21.3~m is shown in figure~\ref{fig:profiles}. The horizontal  axis in the plot is horizontal distance, $y$, from the belt surface and the vertical axis is water surface height above the mean water level.  The profiles are equally spaced in $x$ by 0.012~m and each successive profile is plotted 1.5~mm above the previous profile so that overlap is reduced and the evolution of surface features can be seen.   As in the Couette flow experiment, surface features like ripple crests can be tracked over a number of successive profiles and the slopes of imaginary lines connecting these features are an indication of their horizontal speed away from the belt surface.  It is clear that close to the belt surface, the ripple features last for only a few, say about five, profiles and their paths have a high slope relative to horizontal, indicating relatively slow motion away from the belt.  This region appears to be more chaotic than the region to the right.  In this outer region, surface features remain visible over many frames giving support to the idea that they are freely propagating ripples.  The red line in the figure, which was drawn by eye to approximate that slope of the imaginary lines connection the ripple crests in the outer region, corresponds to a velocity of about 34~cm/s.  It  should be kept in mind that this is only the $y$-component of the phase speed.  Information about the component of the phase speed in the $x$ direction will be obtained in future experiments with the LIF system oriented to record water surface profiles in planes parallel to the belt surface.

As is clear from the profiles in figure~\ref{fig:profiles} and will be seen below in analysis of the RMS surface height fluctuations, the surface motion amplitude is highest close to the belt.    Surface breaking events are more prevalent in this region of high-amplitude surface motion, and, as is discussed below, this appears to be  the primary location for air entrainment. Additionally, due to increased surface roughness and air entrainment at higher belt speeds, automation of surface profile processing is more difficult. Because of this, only data for $U=3.0$~m/s has been fully processed at 1,000 frames per second for 5 runs. For belt speeds of 4.0~m/s and 5.0~m/s, every tenth image was processed, resulting in a time interval of 10~ms between surface profiles. While 5 such runs have been processed for $U=4.0$~m/s, only 2 runs have been processed for $U=5.0$~m/s.

\subsubsection{Surface Fluctuations}

\begin{figure}[!htb]
\begin{center}
\includegraphics[trim=0 0.0in 0 0.00in,clip=true,width=3in]{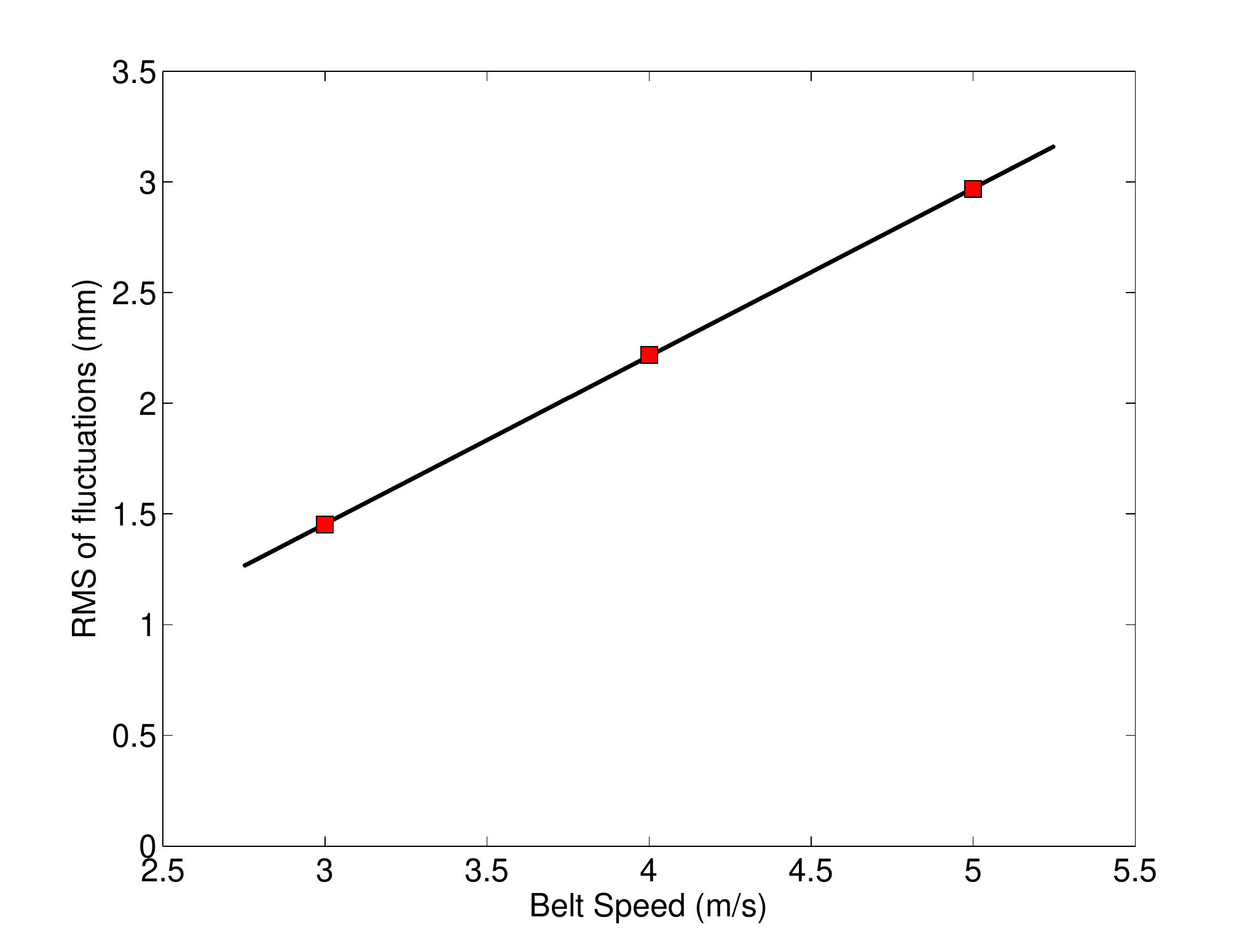}\\
\end{center}
\vspace*{-0.2in} \caption{RMS surface height fluctuation averaged over $y$ and $x$ versus ship speed.} \label{fig:rms_vs_v}
\end{figure} 

\begin{figure*}[!htb]
\begin{center}
\includegraphics[trim=0 0.0in 0 0.00in,clip=true,scale=0.4]{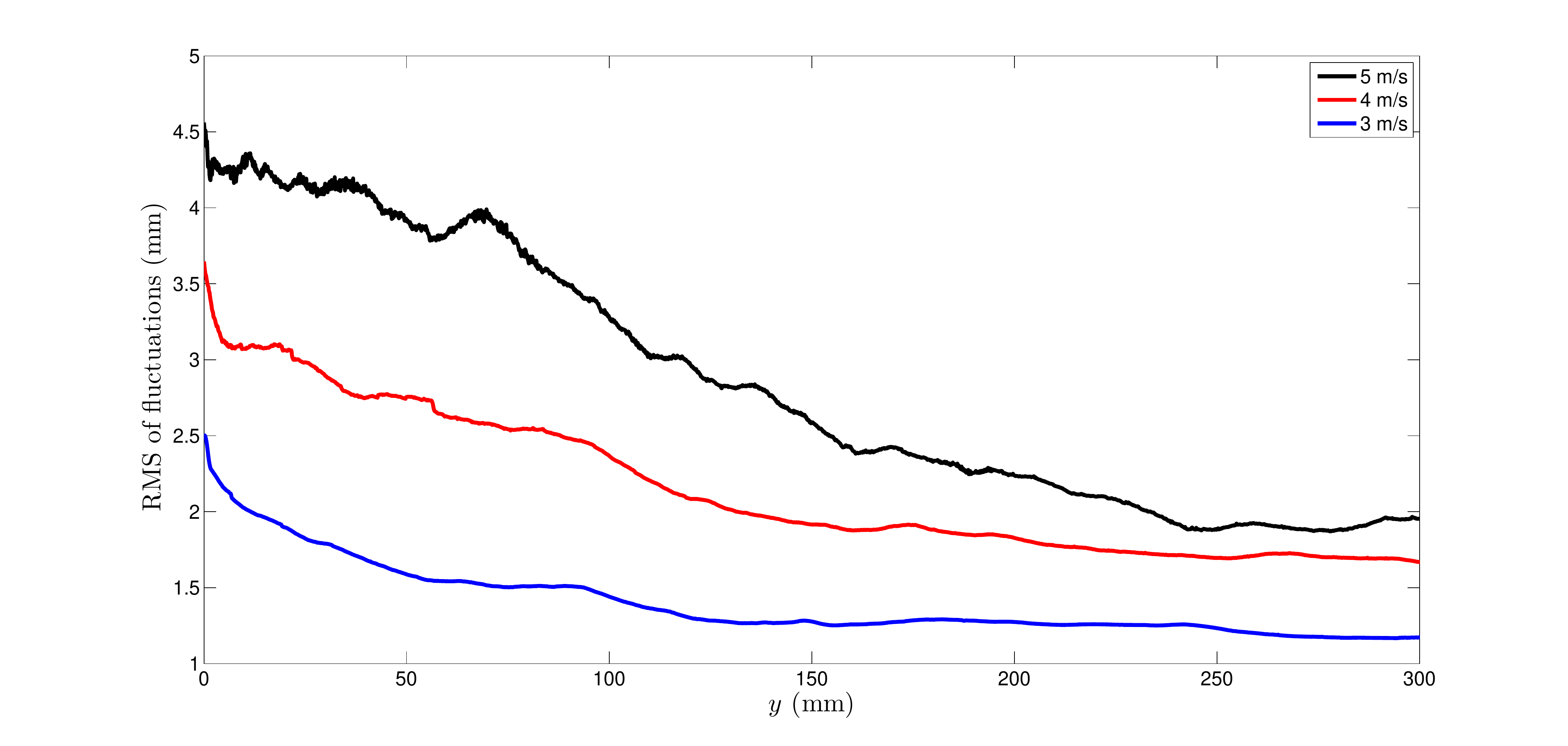}\\
\end{center}
\vspace*{-0.2in} \caption{RMS surface height averaged over  from three to five experimental runs 
versus distance from the belt for three belt speeds.} \label{fig:rms_vs_y}
\end{figure*} 

\begin{figure*}[!htb]
\begin{center}
\includegraphics[trim=0 0.0in 0 0.00in,clip=true,scale=0.4]{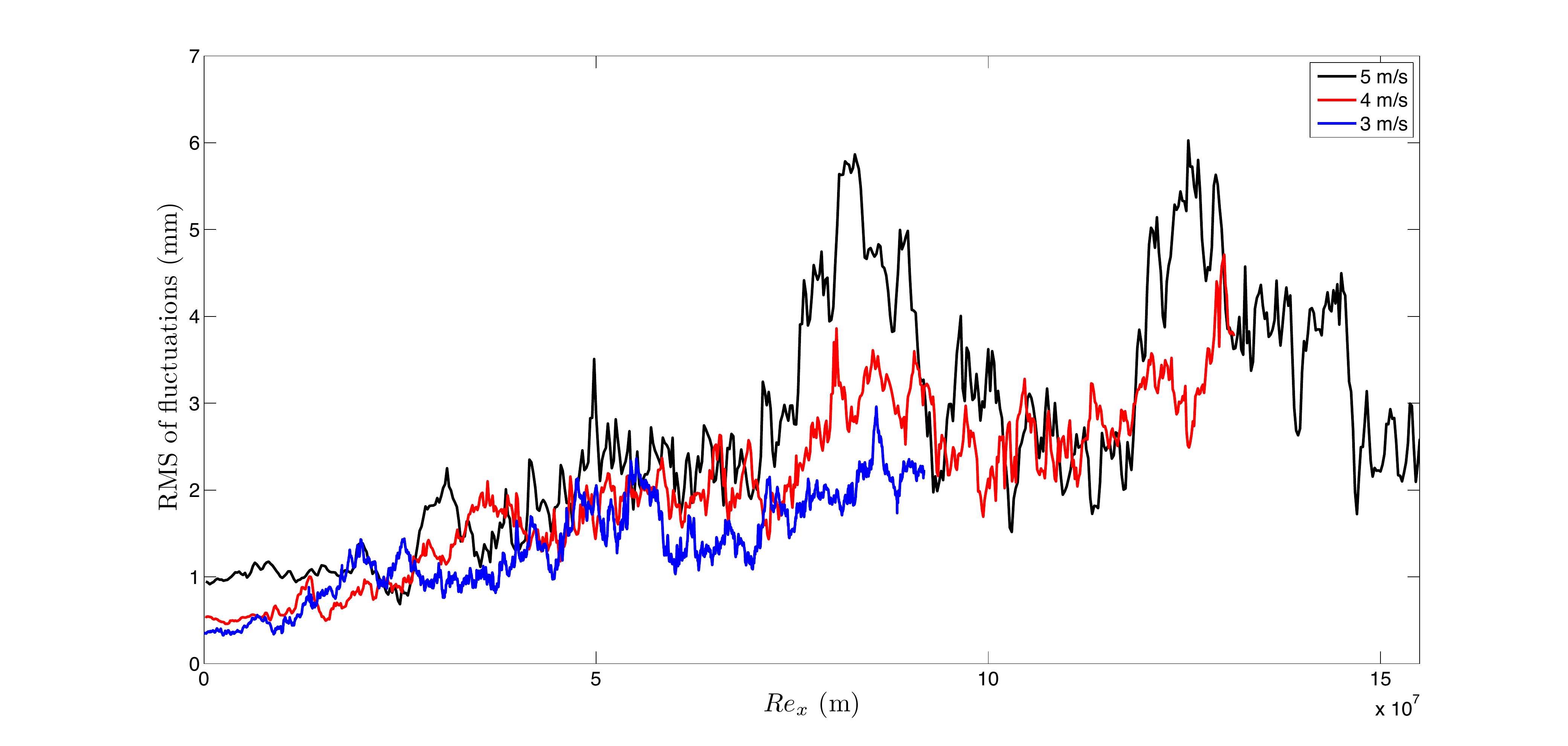}\\
\end{center}
\vspace*{-0.2in} \caption{RMS surface height versus Reynolds number based on distance along the equivalent ship hull ($x=Ut$) for three belt speeds.} \label{fig:rms_vs_x}
\end{figure*} 

In order to characterize these free surface motions, the instantaneous surface profiles are analyzed to determine RMS surface height fluctuations. A mean profile is first determined by averaging the time series of surface height data for each wall-normal position throughout a given run. This average surface height profile is then subtracted from each individual profile and RMS fluctuations about the mean are calculated from the resulting data. (It would be more useful to obtain a temporally evolving mean surface profile by averaging profiles at a given time after the belt begins over many experimental runs.  Unfortunately, this requires a larger data set than was available at the time of writing this paper.)    The RMS surface height is first averaged across both wall-normal position and equivalent ship length to quantify the overall mean fluctuation amplitude for each belt velocity, as shown in figure~\ref{fig:rms_vs_v}. Because these profiles are averaged across every surface height measured for each speed, the RMS values are the most statistically significant of the quantities presented here. The RMS height fluctuations appear to increase linearly with increasing belt speed, but additional data including more belt speeds will be required to gain confidence in this conclusion.

More detail about  the RMS height fluctuation can be found by examining its distributions in $y$ and $x$.  By averaging the RMS over the time series of points for each wall-normal position, the average RMS versus distance from the belt can be determined, as shown in figure~\ref{fig:rms_vs_y}. While the surface fluctuations increase steadily from one speed to the next at all values of $y$, some differences can be seen in the shape of the distributions. Each surface height fluctuation profile reaches a maximum at the belt, but in the highest speed case the values decrease more gradually with increasing distance from the belt than in the lower speed cases. 
In future experiments, turbulent velocity  fluctuations under the water surface will be measured and used to shed light on the reasons for the above-described features in the RMS surface height distributions.  
In order to quantify variation of the RMS surface height fluctuations with distance along the ``hull" ($x$), the RMS fluctuations were averaged over the wall-normal position for each instantaneous surface profile and plotted versus Reynolds number based on equivalent ship length, $x=Ut$, in  figure~\ref{fig:rms_vs_x}. 
As can be seen in the plot, there is a large noise level in the curves.  It was expected that  the curves would show a monotonic increase in surface fluctuation amplitude with increasing $Re_x$ and would collpase together at corresponding $Re_x$ values for different speeds. Unfortunately, the noise level is too high with the present data set to confirm this hypothesis. For a  belt speed of 5~m/s, large spikes in RMS can be seen to reach 2 to 3 times the height of the lower speeds. 

\begin{figure*}[!htb]
\begin{center}
\includegraphics[trim=1.1in -0.5in 0 0.00in,clip=true,scale=0.55]{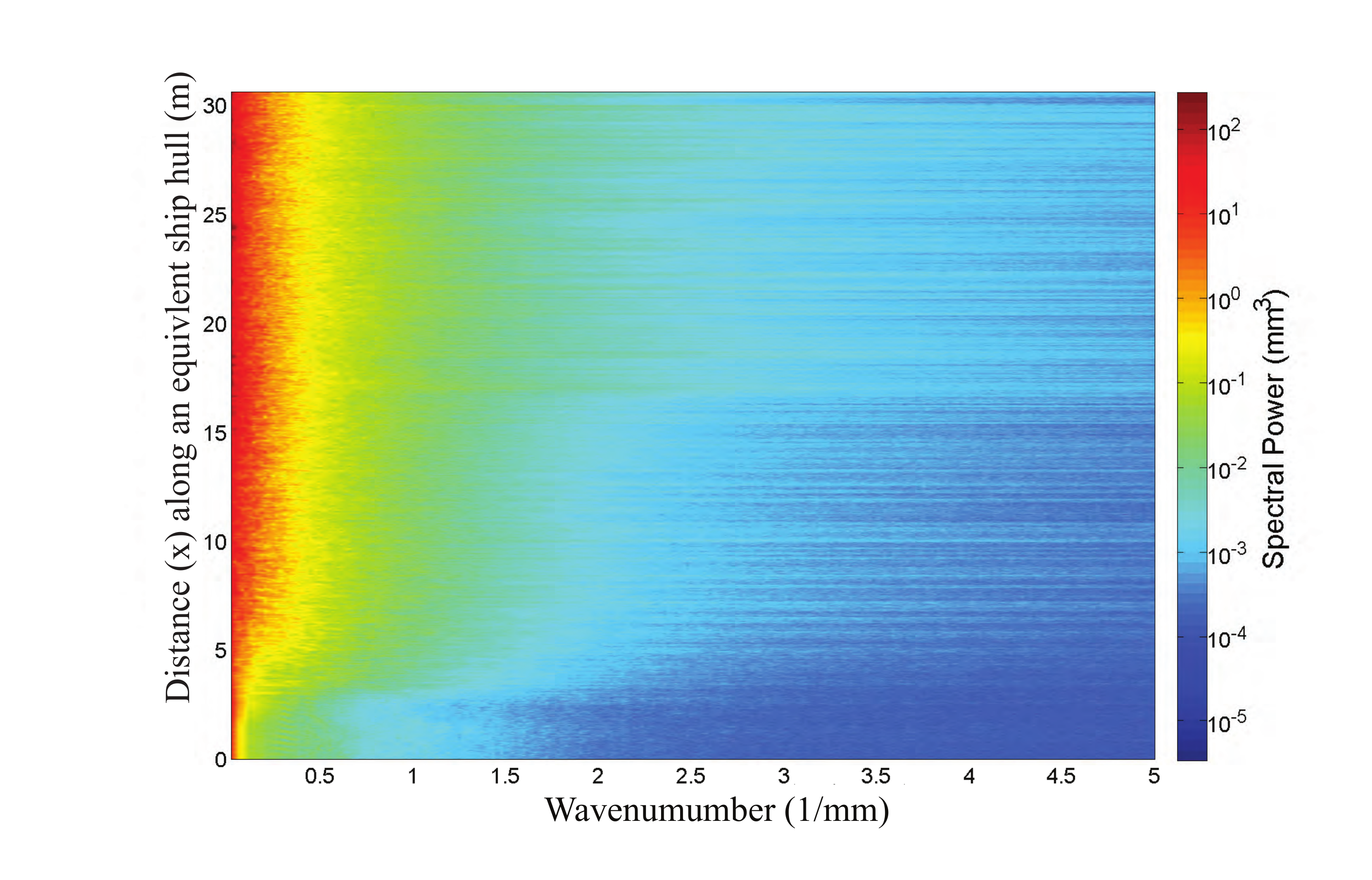}\\
\end{center}
\vspace*{-0.8in} \caption{The spatial fourier transform of the free surface fluctuations was computed for each instantaneous surface profile throughout a run. This plot shows the spectral power vs wavenumber and equivalent ship length using a 20~ms moving average and averaging over 5 runs at 3~m/s.} \label{fig:fft}
\end{figure*} 

The spectral content of the free surface profiles was studied using Fourier transform techniques.  As with the RMS surface fluctuations, a mean surface height for each wall normal position is first subtracted from each profile. Next, a straight line that connects the first and last point is subtracted from each profile. For each resulting instantaneous surface profile, a discrete Fourier transform is calculated, thus converting the signal from the spatial domain $y$ to the wavenumber domain $k$.  
 After repeating this process for each profile, a moving average over 20 profiles is performed for each run and the resulting data from 5 runs is averaged together. In this way, spectral power versus wavenumber and distance along an equivalent ship hull is calculated and shown in figure~\ref{fig:fft}. The spectral content is plotted on a log scale because of the large variation in this quantity.   Before approximately $x=1.5$~m there is very little surface fluctuation  and the spectrum is quite narrow.  For 1.5~m$ <x<$ 5~m, the spectrum grows rapidly in width. This corresponds to a Reynolds number based on equivalent ship length of between $4.5 \times 10^6$ and $15 \times 10^6$.  After this initial sharp increase in surface fluctuations, the spectral shape  change becomes in much more gradual.

\subsubsection{Air Entrainment}

\begin{figure*}[!htb]
\begin{center}
\begin{tabular}{ccc}
(a) & (b)\\
\includegraphics[trim=0 0.0in 0 0.00in,clip=true,scale=0.3]{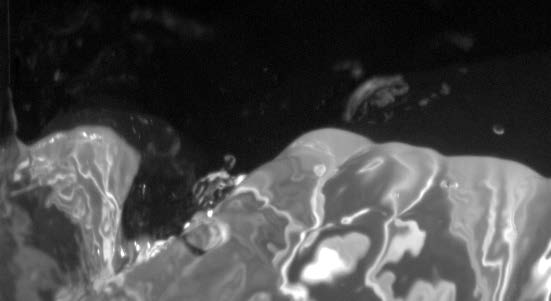}&\includegraphics[trim=0 0.0in 0 0.00in,clip=true,scale=0.3]{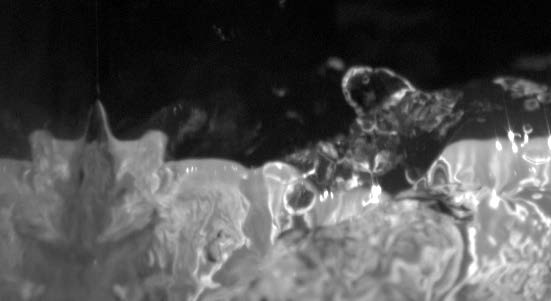}\\
(c) & (d)\\
\includegraphics[trim=0 0.0in 0 0.00in,clip=true,scale=0.3]{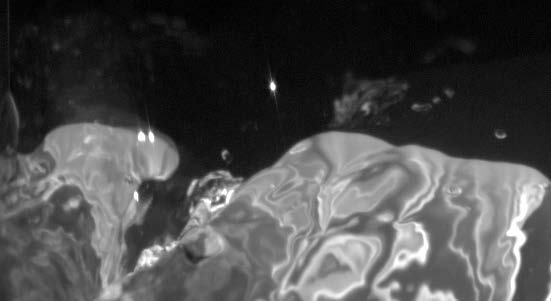}&\includegraphics[trim=0 0.0in 0 0.00in,clip=true,scale=0.3]{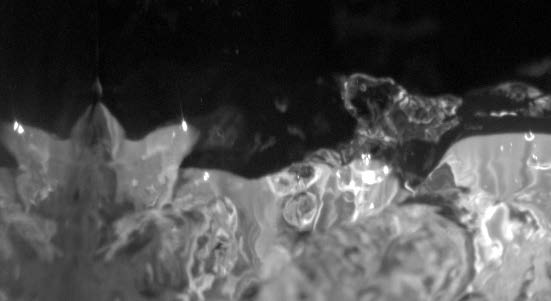}\\
(e) & (f)\\
\includegraphics[trim=0 0.0in 0 0.00in,clip=true,scale=0.3]{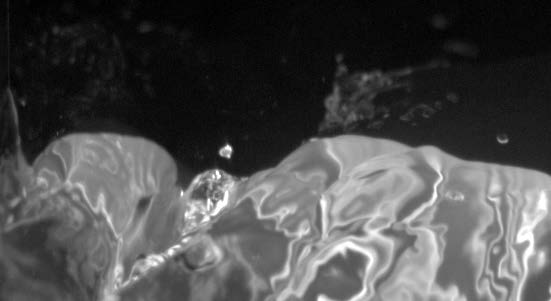}&\includegraphics[trim=0 0.0in 0 0.00in,clip=true,scale=0.3]{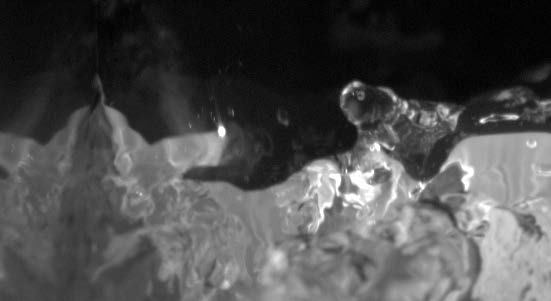}\\
(g) & (h)\\
\includegraphics[trim=0 0.0in 0 0.00in,clip=true,scale=0.3]{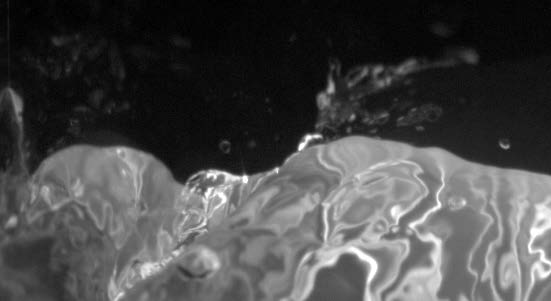}&\includegraphics[trim=0 0.0in 0 0.00in,clip=true,scale=0.3]{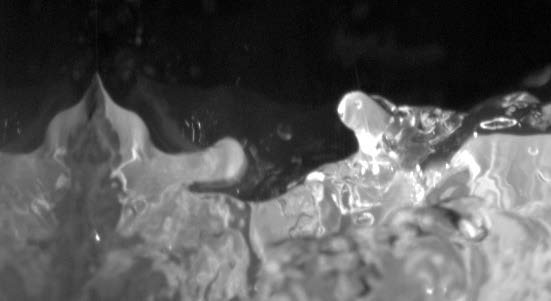}\\
(i) & (j)\\
\includegraphics[trim=0 0.0in 0 0.00in,clip=true,scale=0.3]{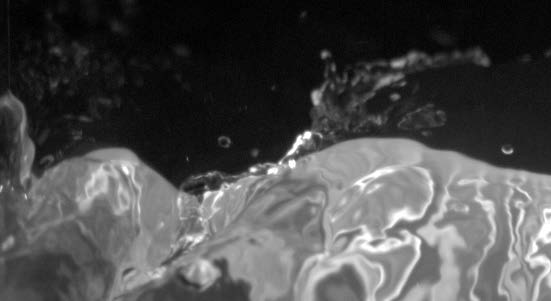}&\includegraphics[trim=0 0.0in 0 0.00in,clip=true,scale=0.3]{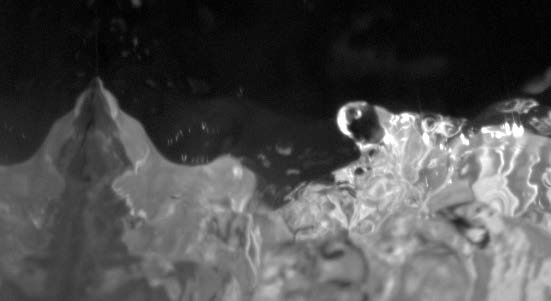}\\
\end{tabular}
\end{center}
\vspace*{-0.2in} \caption{Two sequences of images from LIF movies depicting air entrainment events. Images in the left column depict a trench closure event and images in the right column depict a wave breaking event. Both sets of images are from a belt launch to 5.0~m/s. The horizontal field of view for each image is approximately 7~cm.} \label{fig:waves}
\end{figure*} 

\begin{figure*}[!htb]
\begin{center}
\begin{tabular}{ccc}
(a) & (g)\\
\includegraphics[trim=0 0.0in 0 0.00in,clip=true,scale=0.27]{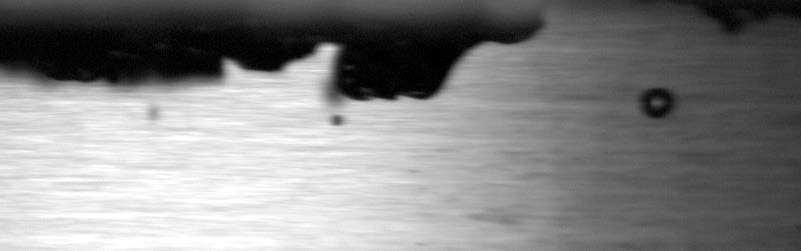}&\includegraphics[trim=0 0.0in 0 0.00in,clip=true,scale=0.27]{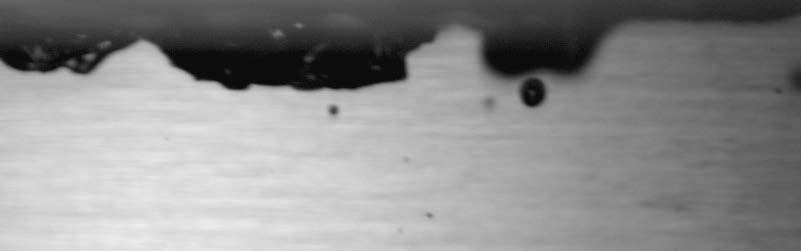}\\
(b) & (h)\\
\includegraphics[trim=0 0.0in 0 0.00in,clip=true,scale=0.27]{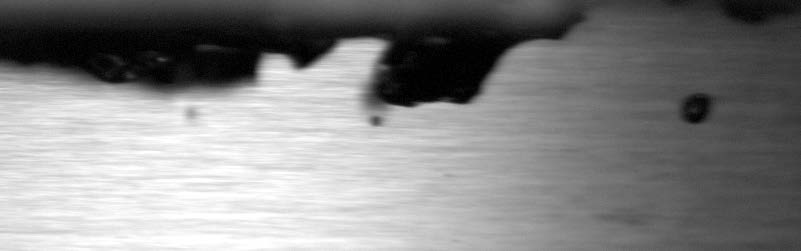}&\includegraphics[trim=0 0.0in 0 0.00in,clip=true,scale=0.27]{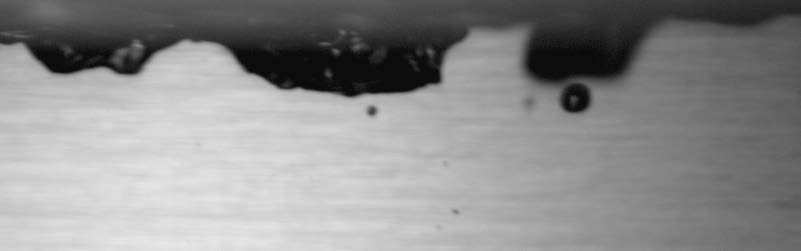}\\
(c) & (i)\\
\includegraphics[trim=0 0.0in 0 0.00in,clip=true,scale=0.27]{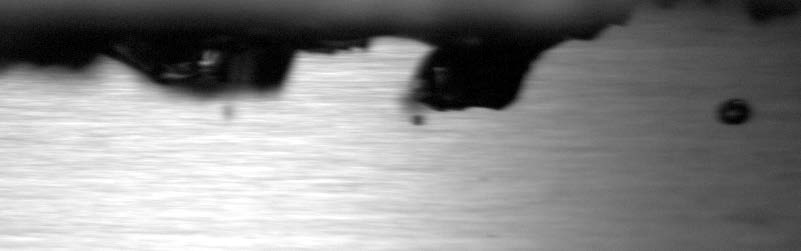}&\includegraphics[trim=0 0.0in 0 0.00in,clip=true,scale=0.27]{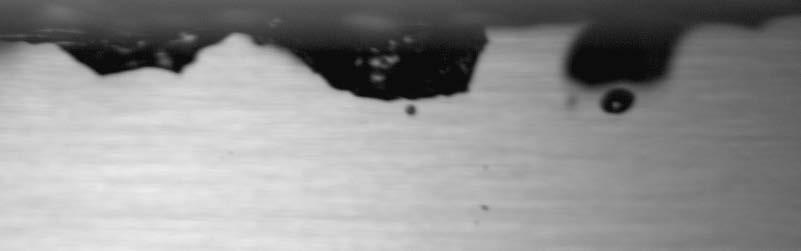}\\
(d) & (j)\\
\includegraphics[trim=0 0.0in 0 0.00in,clip=true,scale=0.27]{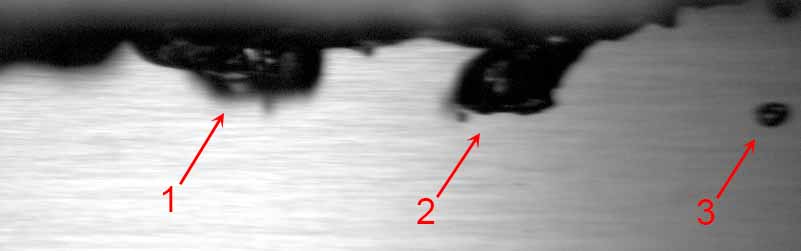}&\includegraphics[trim=0 0.0in 0 0.00in,clip=true,scale=0.27]{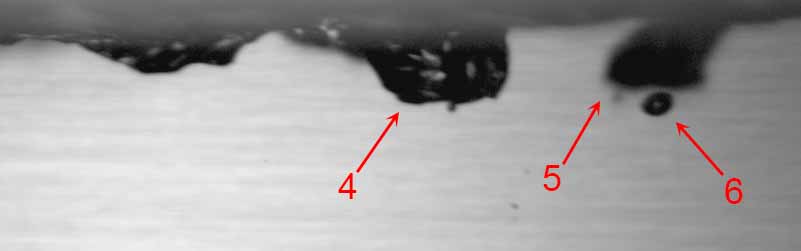}\\
(e) & (k)\\
\includegraphics[trim=0 0.0in 0 0.00in,clip=true,scale=0.27]{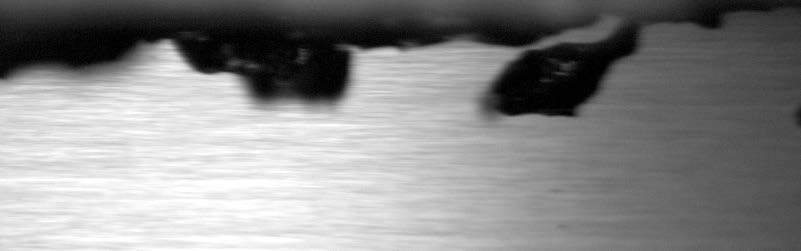}&\includegraphics[trim=0 0.0in 0 0.00in,clip=true,scale=0.27]{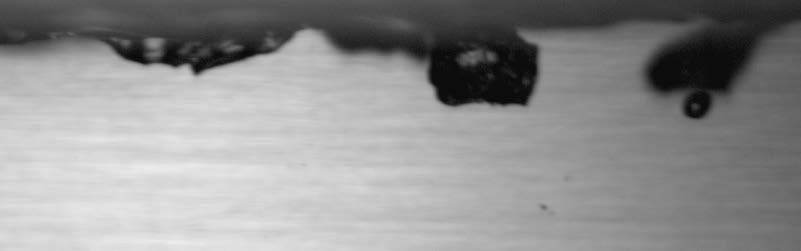}\\
(f) & (l)\\
\includegraphics[trim=0 0.0in 0 0.00in,clip=true,scale=0.27]{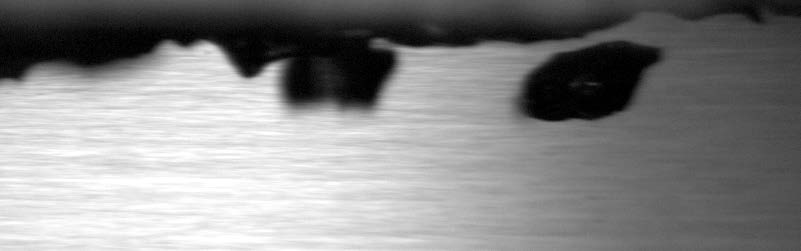}&\includegraphics[trim=0 0.0in 0 0.00in,clip=true,scale=0.27]{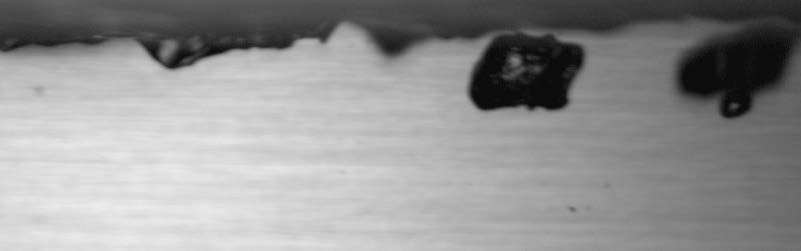}\\
\end{tabular}
\end{center}
\vspace*{-0.2in} \caption{Two sequences of images from underwater white light movies  of the same air entrainment event. The belt is moving from left to right. Images from the left column were captured with the left camera and images from the right camera were captured with the right camera. This event occurred during a launch to 4.0~m/s. Because of the lighting and imaging configuration shown above in figure~\ref{fig:BubbleSchem}, each image contains both the bubble of interest and its shadow on the belt. The horizontal field of view for each image is approximately 6.7~cm.} \label{fig:bubbles_1}
\end{figure*} 

\begin{figure*}[!htb]
\begin{center}
\begin{tabular}{ccc}
(a) & (b)\\
\includegraphics[trim=0 0.0in 0 0.00in,clip=true,scale=0.27]{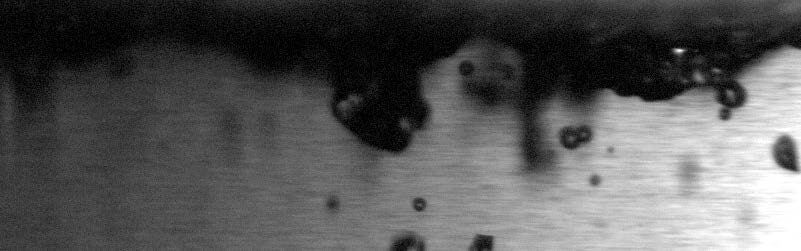}&\includegraphics[trim=0 0.0in 0 0.00in,clip=true,scale=0.27]{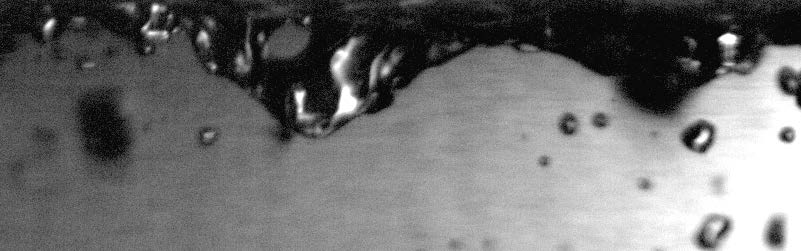}\\
(c) & (d)\\
\includegraphics[trim=0 0.0in 0 0.00in,clip=true,scale=0.27]{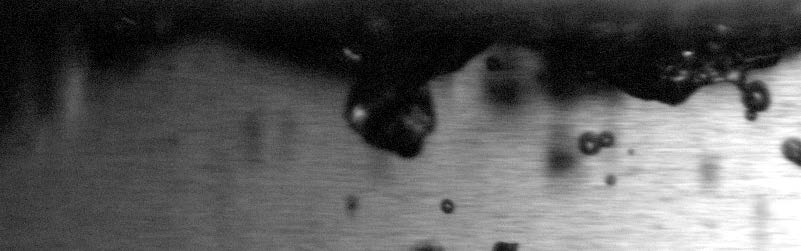}&\includegraphics[trim=0 0.0in 0 0.00in,clip=true,scale=0.27]{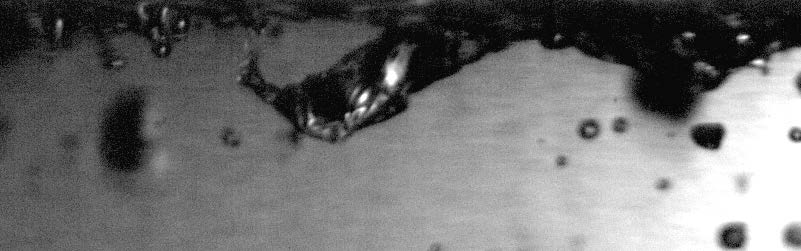}\\
(e) & (f)\\
\includegraphics[trim=0 0.0in 0 0.00in,clip=true,scale=0.27]{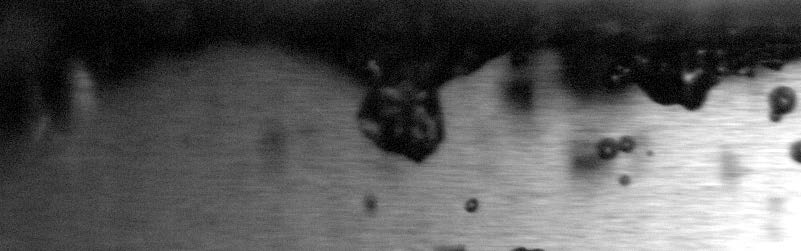}&\includegraphics[trim=0 0.0in 0 0.00in,clip=true,scale=0.27]{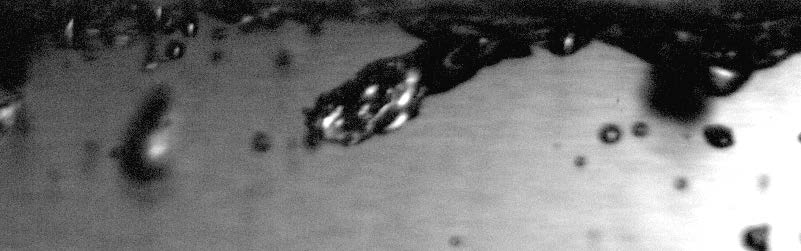}\\
(g) & (h)\\
\includegraphics[trim=0 0.0in 0 0.00in,clip=true,scale=0.27]{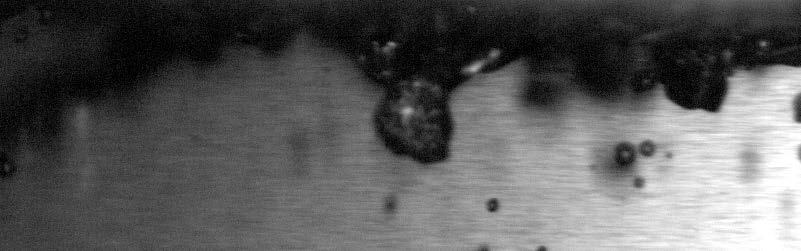}&\includegraphics[trim=0 0.0in 0 0.00in,clip=true,scale=0.27]{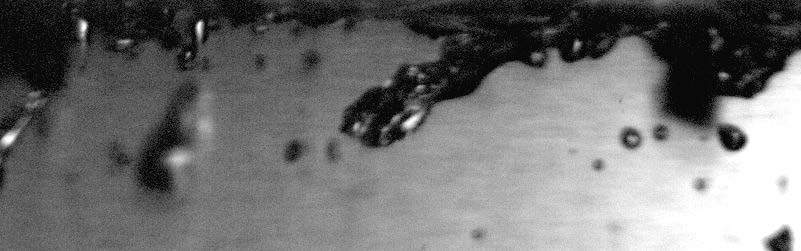}\\
(i) & (j)\\
\includegraphics[trim=0 0.0in 0 0.00in,clip=true,scale=0.27]{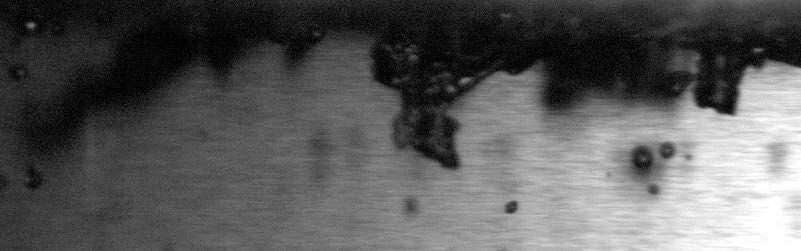}&\includegraphics[trim=0 0.0in 0 0.00in,clip=true,scale=0.27]{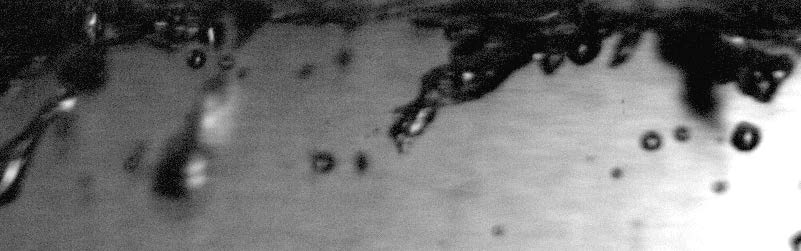}\\
(k) & (l)\\
\includegraphics[trim=0 0.0in 0 0.00in,clip=true,scale=0.27]{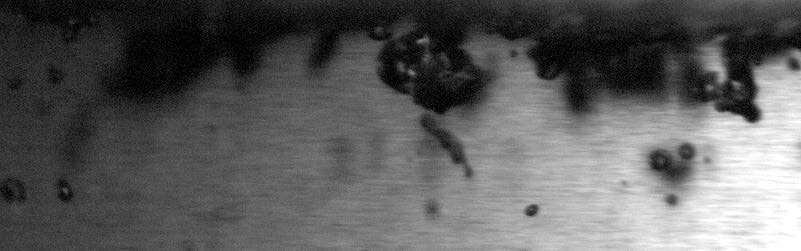}&\includegraphics[trim=0 0.0in 0 0.00in,clip=true,scale=0.27]{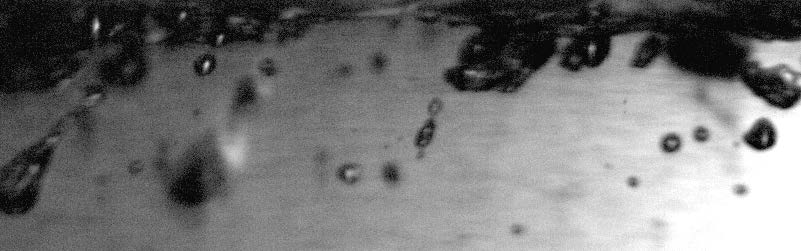}\\
\end{tabular}
\end{center}
\vspace*{-0.2in} \caption{Two sequences of images from underwater white light movies of the same air entrainment event. The belt is moving from left to right. Images from the left column were captured with the left camera and images from the right camera were captured with the right camera. This event occured during a launch to 5.0~m/s. The horizontal field of view for each image is approximately 6.7~cm.} \label{fig:bubbles_2}
\end{figure*} 

From both surface profile movies and underwater white light movies, a variety of air entrainment events can be seen. Perhaps the most prominent of these mechanisms begins with the development of deep narrow troughs that are elongated in a quasi-streamwise direction. Some such troughs become so deep that the sides collapse together. When the sides of the trough meet, air is often entrapped beneath the free surface and bubbles begin to pinch off and convect downstream. Another air entrainment mechanism is caused by ejections from the free surface. These ejections can fold over the surface and trap a pocket of air, similar to the process of air entrainment in plunging breaking waves.

From the LIF movies, some indications of both types of air entrainment events can be seen. Figure~\ref{fig:waves} contains two sets of images to help illustrate the mechanisms for air entrainment as seen from above the free surface. In the left column, a trough collapse is depicted. In images (a) and (c), a deep narrow trough can be seen clearly toward the left side of the image. In these first images, the left side of the trough is beginning to  turn over. In image (e), the two sides of the trough meet and a droplet is ejected into the air. In the right column, the second type of air entrainment event can be seen. In the first image, (b), a jet can be seen forming near the contact point of the free surface with the belt, toward the left side of the image. This jet ejects from the surface and becomes nearly horizontal by image (f). In image (j), the jet makes contact with the free surface, trapping any air that remained below it during the formation of the jet.  Since the camera is above the free surface and because of blockage of the camera line of sight by other free surface features between the camera and the light sheet, the LIF images do not give us a clear indication of whether or not air was actually entrained by either of these particular events.

Another way to visualize these air entrainment events is by recording white light movies beneath the free surface.  This arrangement reduces line of sight blockage by other surface features and allows for a more detailed study of how the air is entrained rather than just showing the surface features that are likely to produce bubbles. Figure~\ref{fig:bubbles_1} shows two sets of images from two views of the same event. Each image contains both the feature of interest and its shadow projected onto the belt by the light source for each camera and can be seen especially well in image pair (d)-(j). In image d, the camera is on the left side, with the light source coming from the right side, so that the actual trough marked as object 2, casts a shadow shown in object 1. In image j, the camera is on the right side with the light source to the left, so the same trough is shown as object 4 with its shadow projected to the right, shown as object 5.  The relative horizontal distance between and object and its shadow is an indication of the distance of the object from the belt, the smaller the distance, the closer the object is to the belt surface.   In the first three sets of images, the trough can be seen to become deep and narrow. In the subsequent images, the left side  of the trough can be seen turning over and contacting the opposite side of the trough, as it pinches off and becomes a large bubble in the final image pair. These events, where one side turns over completely, seem to primarily form large bubbles when entraining pockets of air.

A second type of air entrainment event can be seen in figure~\ref{fig:bubbles_2}. As with the previous figure, this event is shown from two camera views in two columns. This event begins, as shown in image (b) with the two sides of the trough closest and farthest from the right camera meeting at a single point in the middle of the trough and forming a round hole in the trough on the top left of the image.   Following this event, the hole spreads rapidly, propagating outward from the meeting point and forming a large ligament of air that can be seen in image (d). In the following images, this ligament can be seen retracting towards the surface as it deposits a series of small bubbles into the flow.

\begin{figure*}[!htb]
\begin{center}
\includegraphics[trim=0.99in 0 0.0in 0 0.50in,clip=true,scale=0.5]{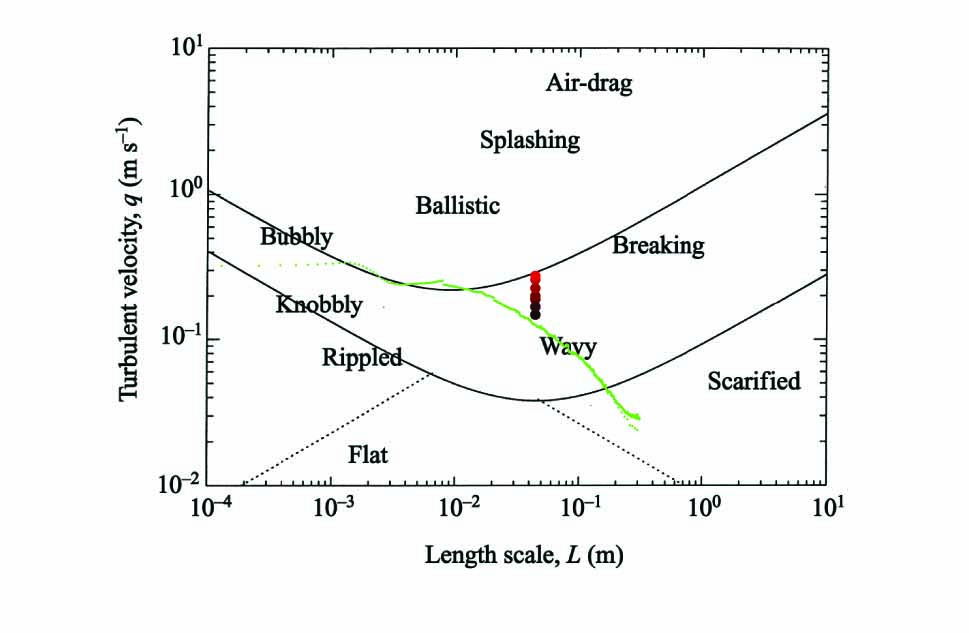}
\end{center}
\vspace*{-0.6in} \caption{Values of RMS fluctuations of $\partial\eta/\partial t$ for both the couette flow and suddenly started belt experiments plotted on the previously discussed chart from \cite{broc:2001} using $\partial\eta/\partial t$ from the experiments in place of the rms vertical velocity fluctuation, $q$, used in the theory.  The large circular data points are from the Couette flow experiments with the highest point corresponding to the highest belt speed and the length scale $L$ taken as the gap width $H$.  The green line is from the suddenly started belt experiment for a belt speed of 3~m/s.     For this curve, the RMS $\partial\eta/\partial t$ data are values at each $y$ averaged over all $x$  and the length scale $L =y$, the distance from the belt surface.   
} \label{fig:brocchiniedit}
\end{figure*}

\subsection{Air Entrainment Boundaries in both the Couette and Suddenly Started Belt Experiments}

The free surface height records at each $y$ location can be differentiated in time to determine the rate of change of surface height ($\partial \eta/\partial t$). In the following, we use this quantity in place of the vertical component of the fluid velocity ($w$) at the surface to compare our results with the entrainment boundary predictions in the theory of \cite{broc:2001} as shown in figure~\ref{fig:brocchiniperegrine}.   By the kinematic boundary condition
\[
w = \frac{\partial \eta}{\partial t} + u\frac{\partial \eta}{\partial x}+v\frac{\partial \eta}{\partial y} \mbox{  on } z = \eta(x,y,t),
\]
thus, the RMS values of $w$ (called $q$ in Brocchini and Peregrine) and $\partial \eta/\partial t$ are not strictly equivalent.  In figure~\ref{fig:brocchiniedit}, we have plotted data from the present experiments  on the original plot of entrainment regions from Brocchini and Peregrine. For the suddenly started belt experiment, only data for $U=3.0$~m/s is presented because this is the only data set for which surface profiles have been processed to a sufficient time resolution to calculate $\partial \eta/\partial t$. 

The points plotted in varying shades of red/black in figure~\ref{fig:brocchiniedit} are from the Couette flow experiments.  For these points, $q$ is taken as the   RMS $\partial\eta/\partial t$ data averaged over the gap width and the length scale $L$ is taken as the gap width, $H$.  As these points transition from black to red, the velocity increases from 2.8~m/s to 4.0~m/s.  
Little or no air entrainment occurred for the belt speeds tested in the Couette flow experiments.  As can be seen in figure~\ref{fig:brocchiniedit}, all of the data points fall on or below the upper curve in the plot.  According to the theory, for flows that correspond to points above this boundary air entrainment may occur.  Note that the position of the data points may indicate that a slightly higher belt speed will result in air entrainment.  

The green line in figure~\ref{fig:brocchiniedit} is data from the suddenly started belt experiments.  The RMS $\partial\eta/\partial t$ data are values at each $y$ averaged over all $x$  and the length scale $L$ is taken as $y$, the distance from the belt surface.   
The data points are on or below the air entrainment boundary  with the exception of the points in the range 
between $y=L=0.1$~cm and about $1.5$~cm, which are slightly above the boundary.  In the experiment, $U=3.0$~m/s is just below the entrainment boundary, in approximate agreement with the Brocchini and Peregrine prediction.  In future experiments, the stereo imagery described above will be used to determine the range of $y$ where the air entrainment occurs.

\section{4. Conclusions}

A laboratory-scale device is used to study the interaction of turbulent boundary layer shear flows with a free surface, including air entrainment.  Two experiments are performed.  In the first experiment, a stationary flat plate is placed parallel to and at a distance $H$ from a vertical  wall that moves horizontally in its own plane.  Both the moving wall and the fixed plate pierce the water surface. The wall is run at constant speed until a steady-state Couette flow is created. In the second experiment, the wall motion consists of 3$g$ acceleration followed by a steady speed, $U$.  In this case, the fixed plate is absent and the flow represents a temporally evolving free surface boundary layer.  Via the transformation $x=Ut$, this experiment mimics the boundary layer on a surface piercing plate moving horizontal at constant speed.   Water surface profiles are recorded with a cinematic LIF system in both experiments to study the generation of surface height fluctuations by the subsurface turbulence. Stereo, underwater white-light movies are recorded in the suddenly started belt experiment to determine the mechanisms for air entrainment. In the Couette flow experiment, it is found that surface fluctuations increase sharply near both the moving belt and stationary wall and that there is a peak in the  spectra of the rate of change of surface height at a non-dimensional frequency ($fH/U$) of about 0.7.  In the suddenly started belt experiment, it is found that after a relatively calm beginning portion of each experimental run, surface height fluctuations increase sharply before leveling off later in the run. From underwater white-light movies, different types of air entrainment events are detected at small distances away from the belt, entraining both large and small bubbles. Further experiments are planned  to increase the statistical significance of measured quantities, to extend the range of wall speeds, obtain quantitative data from the stereo bubble entrainment movies  and to obtain subsurface flow field data using cinematic, stereo particle image techniques. 
\section{Acknowledgements}

The authors gratefully acknowledge the support of the Office of Naval Research under grant N000141110029,  Program Manger Dr.\ Ki-Han Kim.

\bibliography{29ONR}
\bibliographystyle{29ONR}

\end{document}